\documentclass[a4paper]{spie}

\usepackage{amsmath,amsfonts,amssymb}
\usepackage{graphicx}
\usepackage[colorlinks=true, allcolors=blue]{hyperref}
\usepackage{xspace}
\usepackage{txfonts}
\usepackage{subcaption}

\newcommand{\teff}{\ensuremath{\mathrm{T}_{e\!f\!f}}\xspace}
\newcommand{\logg}{\ensuremath{\log g}\xspace}

\newcommand{\mic}{$\mu$m\xspace}
\newcommand{\as}{\hbox{$^{\prime\prime}$}\xspace}

\title{Bringing high-spectral resolution to VLT/SPHERE with a fibre coupling to VLT/CRIRES+}

\author[a]{A. Vigan}
\author[a]{G. P. P. L. Otten}
\author[a]{E. Muslimov}
\author[a]{K. Dohlen}
\author[b]{M. W. Phillips}
\author[c]{U. Seemann}
\author[a]{J.-L. Beuzit}
\author[d]{R. Dorn}
\author[d]{M. Kasper}
\author[e]{D. Mouillet}
\author[b]{I. Baraffe}
\author[c]{A. Reiners}

\affil[a]{Aix Marseille Univ, CNRS, CNES, LAM, Marseille, France}
\affil[b]{Astrophysics Group, University of Exeter, Exeter, EX4 4QL, UK}
\affil[c]{Institut f\"ur Astrophysik, Georg-August-Universit\"at, Friedrich-Hund-Platz 1, D-37077 G\"ottingen, Germany}
\affil[d]{European Southern Observatory (ESO), Karl-Schwarzschild-Str. 2, 85748 Garching, Germany}
\affil[e]{Univ. Grenoble Alpes, CNRS, IPAG, 38000 Grenoble, France}

\authorinfo{Send correspondence to A. Vigan: arthur.vigan@lam.fr}

\begin{document}
\maketitle

\begin{abstract}
Atmospheric composition provides essential markers of the most fundamental properties of giant exoplanets, such as their formation mechanism or internal structure. New-generation exoplanet imagers, like VLT/SPHERE or Gemini/GPI, have been designed to achieve very high contrast ($>15$ mag) at small angular separations ($<$0.5\as) for the detection of young giant planets in the near-infrared, but they only provide very low spectral resolutions ($R<100$) for their characterization. High-dispersion spectroscopy at resolutions up to $10^5$ is one of the most promising pathways for the detailed characterization of exoplanets, but it is currently out of reach for most directly imaged exoplanets because current high-dispersion spectrographs in the near-infrared lack coronagraphs to attenuate the stellar signal and the spatial resolution necessary to resolve the planet. Project HiRISE (High-Resolution Imaging and Spectroscopy of Exoplanets) ambitions to develop a demonstrator that will combine the capabilities of two flagship instruments installed on the ESO Very Large Telescope, the high-contrast exoplanet imager SPHERE and the high-resolution spectrograph CRIRES+, with the goal of answering fundamental questions on the formation, composition and evolution of young planets. In this work, we will present the project, the first set of realistic simulations and the preliminary design of the fiber injection unit that will be implemented in SPHERE.
\end{abstract}



\section{Introduction}
\label{sec:introduction}

The new generation of exoplanet imagers now available on 8-10~m-class telescopes\cite{Jovanovic2015,Macintosh2014,Beuzit2008} offers significant perspectives in terms of young giant exoplanet discovery\cite{Macintosh2015,Chauvin2017} and characterisation\cite{Zurlo2016,Vigan2016a,Samland2017,Chilcote2017,Greenbaum2018}. However, despite the major gain offered by extreme adaptive optics (ExAO) systems and efficient coronagraps\cite{Soummer2005,Boccaletti2008,Guyon2014}, the data remain limited by variable AO residuals and quasi-static speckles\cite{Soummer2007,Vigan2016b}. The use of clever observing strategies\cite{Racine1999,Marois2006a} and advanced data analysis methods\cite{Marois2006a,Lafreniere2007,Mugnier2009,Soummer2012,Cantalloube2015} partly alleviates these systematics, but in most cases the final accuracy of the astrometry or photometry on planetary companions remains limited by the speckle residuals\cite{Zurlo2016,Wang2016,Nielsen2017,Wertz2017,Samland2017}. These limitations make the direct spectral characterisation of planetary-mass companions challenging when their contrast becomes higher than $10^5$ and their separation smaller than 1\as. In addition, most new-generation instruments have been designed to include integral-field spectrographs (IFS) that provide simultaneous spatial and spectral information, but only at relatively low spectral resolution (typically R$<$100 in GPI or SPHERE). These measurements can be used to constrain the basic atmospheric properties of planets, like effective temperature or surface gravity, but their very low resolution leads to degeneracies when fitting physical models\cite{Bonnefoy2016} and does not allow focused measurements like abundances and rotational period determination.

Another pathway for enabling advanced characterization of exoplanets is to use high-dispersion spectroscopy 
with resolving powers of the order of several $10^4$ or even $10^5$. 
These resolving powers enable to unambiguously disentangle the spectral features of the star and the planet, 
especially when considering that the planet has a distinct radial velocity component originating from its orbital motion. This technique has already been demonstrated to work in a variety of configurations, in the near-infrared and in the visible, to study the orbital motion, mass, high-altitude winds, composition, or albedo of a handful of exoplanets\cite{Snellen2010,Crossfield2012,Martins2015}. Nonetheless, the capacity to detect the faint signal of the planet embedded in the signal of the star remains limited by the difference in the luminosity (i.e. the contrast) between the two objects\cite{Hoeijmakers2018}, which clearly calls for other techniques to increase the contrast accessible in the vicinity of bright stars.

The idea of combining high-contrast imaging with high-dispersion spectroscopy to enable the detection of extremely faint planets has already been proposed several times in the past\cite{Sparks2002,Riaud2007,Snellen2015}, and in fact self-luminous, young, giant exoplanets constitute ideal targets because of their intrinsic brightness in the near-infrared. The characterization of $\beta$\,Pictoris\,b\cite{Lagrange2010} using VLT/CRIRES at a resolution of $10^5$ in $K$-band\cite{Snellen2015} and of HR\,8799\,c using Keck/OSIRIS at R=4000 in $K$-band\cite{Konopacky2013} has clearly demonstrated the potential of the technique on these objects, but so far other directly imaged exoplanets remain out of reach because current high-dispersion spectrographs in the near-infrared lack coronagraphs to attenuate the stellar signal and the spatial resolution necessary to properly resolve the planets. The commissioning of new-generation exoplanet imagers\cite{Beuzit2008,Macintosh2014,Jovanovic2015} or the definition of an upgrade path for other existing instruments\cite{Mawet2016} has spurred recent studies that propose the coupling of high-contrast imaging instruments with high-resolution spectrographs, with the specific goal of bringing exoplanet detection and characterisation to the next level\cite{Wang2017,Mawet2017,Lovis2017}.

In this paper, we present the current status of project HiRISE (High-Resolution Imaging and Spectroscopy of Exoplanets) at the ESO Very Large Telescope, which aims at implementing a near-infrared fibre coupling between SPHERE and CRIRES+ to enable the characterisation of young giant exoplanets at high-spectral resolution. In Section~\ref{sec:general_implementation} we first briefly discuss the general implementation at VLT/UT3. Then in Section~\ref{sec:fiu_sphere} we describe the SPHERE infrastructure and the proposed design of a fibre injection unit. In Section~\ref{sec:crires+} we provide a short update on the CRIRES+ instrument and how a fibre bundle coming from SPHERE could be implemented. Finally, in Section~\ref{sec:simulations}, we present the first performance simulations of the system before concluding in Section~\ref{sec:conclusion_prospects}.

\section{General implementation at the telescope}
\label{sec:general_implementation}

\begin{figure}
  \centering
  \includegraphics[width=1\textwidth]{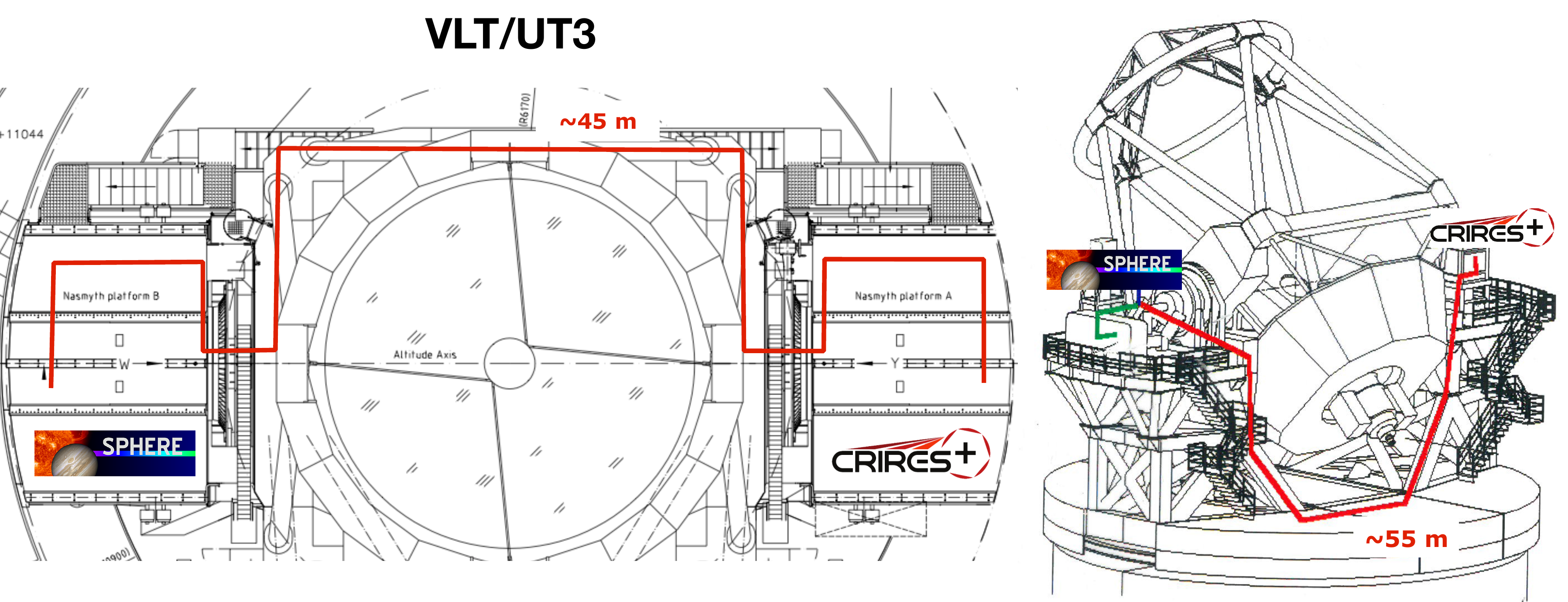}
  \caption{Two possible solutions for the implementation of a fibre between the two VLT Nasmyth platforms: either going along the primary mirror cell (left) or going down betneath the azimuth floor and back up (right). In both cases the total length of fibre is of the order of $\sim$50~m. The option on the right has already been implemented in Paranal for the FLAMES instrument\cite{Pasquini2002}, which hosts a fibre positioner that can be used to feed the UVES spectrographs.}
  \label{fig:fibre_path}
\end{figure}

The general implementation of this coupling is based on window of opportunity at the ESO Very Large Telescope in Chile, where the UT3 will soon host two flagship instruments: on one side the SPHERE exoplanet imager\cite{Beuzit2008} designed for very high-contrast imaging in the near-infrared, and on the other side the CRIRES+ very high-resolution spectrograph\cite{Dorn2016}, which covers the same spectral bands as SPHERE but includes neither extreme adaptive optics nor coronagraphy. This unique setup provides an opportunity to explore the combination of two techniques that may well be the future of exoplanet direct imaging and characterisation\cite{Snellen2015,Wang2017}, while being able to produce high-level scientific results that are currently unreachable by either SPHERE or CRIRES+ on their own.

The practical implementation of the coupling requires 3 separate elements:

\begin{enumerate}
    \setlength{\itemsep}{1pt}
    \setlength{\parskip}{0pt}
    \setlength{\parsep}{0pt}
    \item\label{it:fiu} an opto-mechanical interface in SPHERE that enables to pick-up the signal of known directly imaged exoplanets in SPHERE, which we will refer to as a \emph{fibre injection unit}\cite{Mawet2017};
    \item\label{it:fibre} an near-infrared fibre bundle;
    \item\label{it:feu} an opto-mechanical interface in CRIRES+ that aligns the input fibres along the slit of the spectrograph and adapts the beam focal ratio to the value expected by the instrument.
\end{enumerate}

\noindent The \ref{it:fiu}$^{\mathrm{st}}$ element will be described in detail in Section~\ref{sec:fiu_sphere} and the \ref{it:feu}$^{\mathrm{rd}}$ will be briefly mentioned in Section~\ref{sec:crires+}. The \ref{it:fibre}$^{\mathrm{nd}}$ element needs some clarification because, although the principle of the implementation is simple, the actual implementation is not straightforward because of the massive size of the VLT/UT3 and therefore the distance between the two instruments. Two possible options are pictured in Figure~\ref{fig:fibre_path}. The first option, which requires the shortest length of fibre (40--45~m), would be to go from one Nasmyth platform to the other by following a path around the cell of the primary mirror. However, this options probably requires a cable wrap system because the mirror cell is rotating around the altitude axis during the observations. The second option goes down the telescope structure, beneath the azimuth floor and then back up to the other Nasmyth platform. While this option requires a slightly longer fibre length ($\sim$55~m), it can be completely static because access to azimuth floor is part of the telescope structure. In fact, this second option has already been implemented for the FLAME instrument\cite{Pasquini2002}, which hosts a fibre positioner that can be used to feed the UVES spectrograph. It is likely that the HiRISE project will adopt a similar setup.

\section{Fibre injection unit for SPHERE}
\label{sec:fiu_sphere}

\subsection{The SPHERE infrastructure}
\label{sec:sphere_infrastructure}

The SPHERE planet finder instrument\cite{Beuzit2008} installed at the VLT is a highly specialized instrument, dedicated to high-contrast imaging and spectroscopy of young giant exoplanets. It is constituted of a common path interface (CPI) that provides extremely high-quality optics, such as stress polished toric mirrors\cite{Hugot2012}, to transport the beam from the entrance of the instrument down to the coronagraphs and scientific instruments. At the heart of the instrument is the SAXO ExAO system\cite{Fusco2006,Petit2014,Sauvage2014}, based on a spatially-filtered Shack-Hartmann wavefront sensor (SHWFS), which controls a $41\times41$ actuators deformable mirror at a frequency up to 1.3\,kHz. SAXO controls four loops for the stabilisation of the fast visible tip-tilt, the high-order modes, the near-infrared differential tip-tilt, and the pupil. The fast tip-tilt, high-order and pupil stabilization loops benefit all the scientific instruments, while the near-infrared differential tip-tilt loop obviously applies only to the two NIR instruments. 

Although SPHERE was designed to provide good turbulence correction in the visible for science with the ZIMPOL instrument\cite{Thalmann2008}, its primary technical requirement was to provide diffraction-limited images in the near-infrared to search for self-emitting, young, giant exoplanets. Looking for these planets in the near-infrared requires a careful trade-off between different constraints on the wavelength: accessing small angular separations favors small wavelengths (because of the diffraction size), but the expected contrast between stars and young planets tends to be for favorable towards longer wavelengths. In addition, the Strehl ratio (directly related to the turbulence correction) improves with longer wavelengths, at the expense of the thermal background from the instrument and the sky, which increase with wavelength (especially in $K$-band and upwards), requiring to cool down the instrument. In SPHERE, a compromise was found by enabling simultaneous observations from 0.95 to 2.3~\mic with two science instruments observing in parallel, IFS\cite{Claudi2008} and IRDIS\cite{Dohlen2008}. These simultaneous observations require the use of either sufficiently broad-band coronagraphs, such as an achromatic four-quadrants phase masks\cite{Boccaletti2008}, or a coronagraph optimised for the longest wavelength access in the observations, such as the apodized pupil Lyot coronagraphs\cite{Soummer2005}, which is the baseline of current SPHERE observations.

Most small-inner working angle coronagraphs are very sensitive to low-order aberrations such as tip, tilt, focus and the few higher-order aberrations. In SPHERE, the tip, tilt and focus are directly optimised at the level of the coronagraph during the target acquisition sequence, but the higher order non-common path aberrations are currently not compensated\cite{Vigan2018}. In order to minimise the effect of residual tip-tilt jitter at the level of the coronagraph, SPHERE uses a differential tip-tilt sensor\cite{Baudoz2010} (DTTS). This sensor picks up a very small fraction ($\sim$1\%) of the near-infrared beam just before the coronagraph and produces a non-occulted image of the star that is used by the DTTS loop to stabilize the PSF on the focal-plane mask of the coronagraph, with a typical accuracy of a few milliarcseconds. After the coronagraph, the near-infrared beam is split with a dichroic that sends the short wavelengths ($<$1.7~\mic) to the IFS arm and the long wavelengths ($>$1.7~\mic) to IRDIS. Each of the two instruments has its own Lyot stop: in the cryostat for IRDIS, and on the CPI for the IFS. 

Downstream of the coronagraphic mask and the Lyot stop, the near-infrared beam available in SPHERE benefits from absolutely all the features required for high-contrast imaging of exoplanets: stable environment, high-quality optics, high-order AO correction, very fine stabilisation of the tip-tilt, and full-fledged coronagraph.

\subsection{Fiber injection unit location and conceptual design}
\label{sec:location_conceptual_design}

\begin{figure}
  \centering
  \includegraphics[width=1\textwidth]{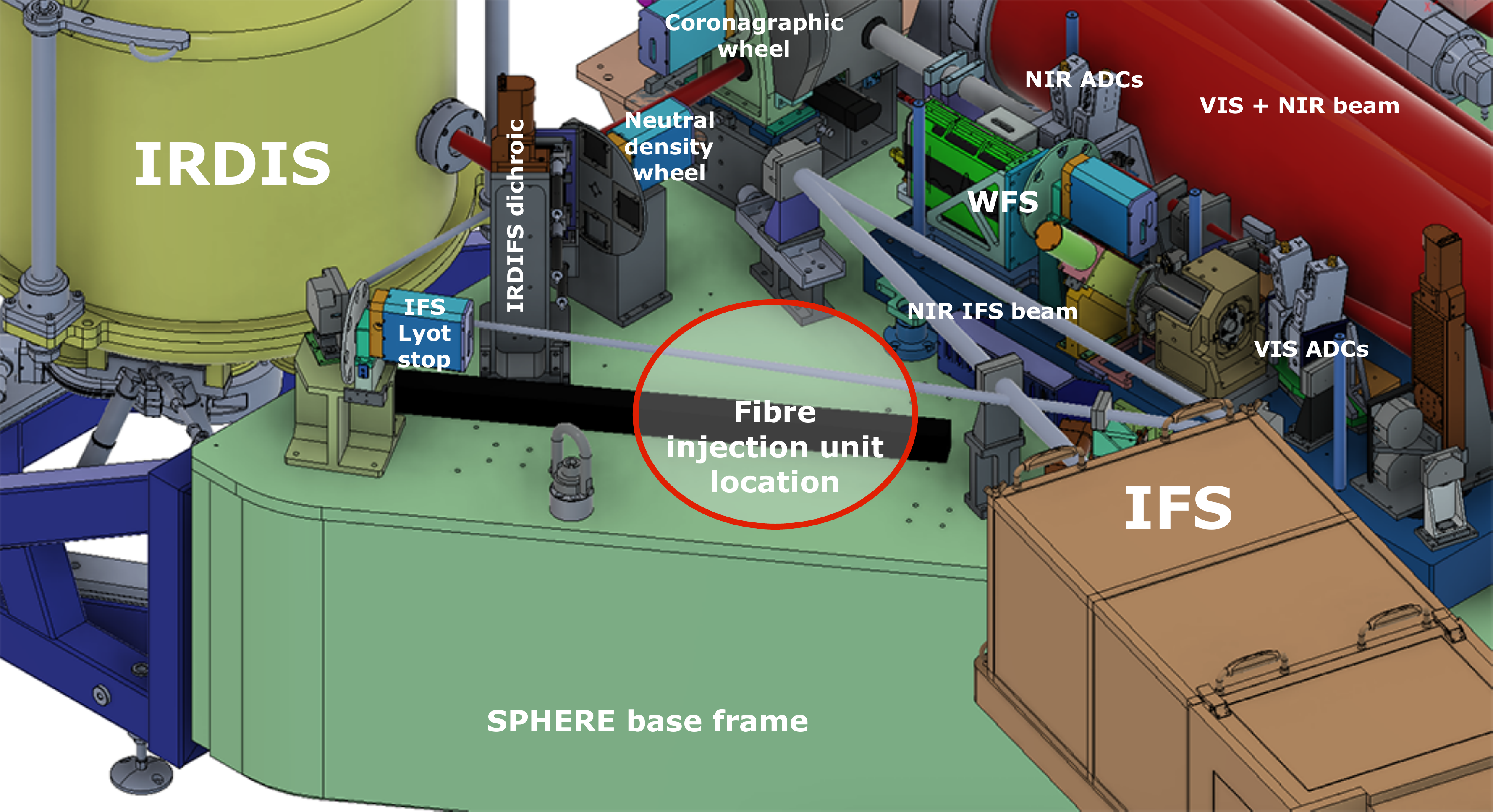}
  \caption{Foreseen location for the fibre injection unit in SPHERE. This location benefits from the complete SPHERE infrastructure: high-quality optics, stable environment, ExAO, second-stage tip-tilt stabilisation and full APLC coronagraph. In addition, this location offers some free space on the SPHERE base frame where the unit can be easily implemented without impact on the normal operations of the instrument.}
  \label{fig:hirise_cpi_location}
\end{figure}

The are a few options available for the implementation of a fibre injection unit in SPHERE. Since this implementation requires a focal-plane, a first option could be to use the existing focal plane located at the level of the coronagraphic wheel, which contains the focal-plane masks of the various coronagraphs. However, this location is packed with elements (NIR beam, coronagraphic wheel and its motor, DTTS pick-off and cryostat) so it is difficult to foresee how to fit a fibre bundle plus the necessary hardware to position it accurately. Picking off the beam shortly after the coronagraphic mask is also impractical because of the neutral density wheel (and its motor) immediately followed by the IRDIFS dichroic beam splitter. In addition, none of the above options are advantageous in terms of coronagraphy because they are located before the Lyot stops, which would prevent benefiting from the full diffraction attenuation unless a dedicated Lyot stop or pupil apodiser is used. 

To benefit from the full attenuation of the coronagraph, the most convenient location is actually in the IFS arm (see Figure~\ref{fig:hirise_cpi_location}). At this location, the NIR beam is almost collimated and extended over a significant length, with two folding mirrors, before it enters the IFS. In addition, this area on the CPI base frame is mostly free of any opto-mechanical elements, which provides significant space, and it is easily accessible from the outside because the SPHERE enclosure provides an access door just in front of this area. It is therefore the perfect location to implement an additional sub-system in SPHERE.

\begin{figure}
  \centering
  \includegraphics[width=0.65\textwidth]{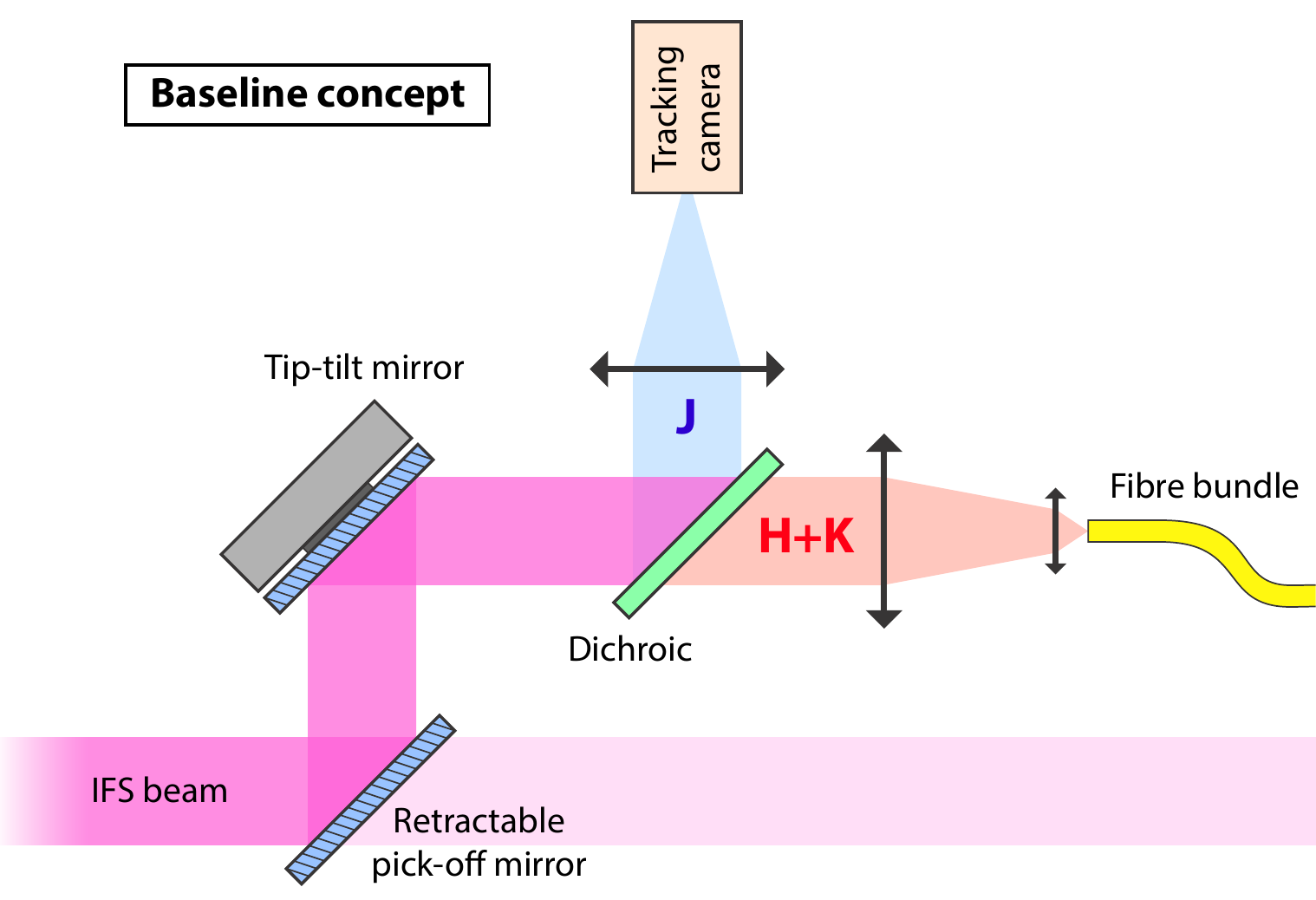}
  \caption{Conceptual sketch of the SPHERE fibre injection unit located in the IFS arm. The IFS beam is picked off using a retractable mirror. The unit has two arms separated by a dichroic beam splitter that sends the $J$-band to a tracking camera and the $HK$-bands to a single-mode fibres bundle. A tip-tilt mirror enables to move the image with respect to the tracking camera and the fibres.}
  \label{fig:hirise_fiu_concept}
\end{figure}

The proposed conceptual design of the SPHERE fibre injection unit is shown in Figure~\ref{fig:hirise_fiu_concept}. In order to avoid any harmful interaction with the rest of the system, the FIU is designed as a completely independent system. It will include a retractable pick-up mirror that can be inserted into the IFS beam to take all the light, i.e. the IFS will not receive any light when the FIU is in use. The instrument pupil will then be re-imaged on a controllable tip-tilt mirror, which will be used to move the focal plane image with respect to the fibre bundle and tracking camera. After the tip-tilt mirror, a dichroic plate will split the beam to send the short wavelengths ($Y$- and $J$-band) to a tracking camera and the long wavelengths ($H$- and $K$-band) to the fibre bundle. 

The layout of the fibres in the fibre bundle is not completely defined yet. In addition to the science fibre, which will pick up the light of the planet, there will be at least 1 reference fibre to pick up the light from the star (at some location in the speckle field), but in fact several reference fibres can be foreseen to make sure that at least one of them always samples the speckles in an area where sufficient stellar flux is available. An additional centring fibre is currently foreseen (see Section~\ref{sec:operational_scheme}) for the calibration of the interaction matrix of the tip-tilt mirror and the fine positioning of the planet on the science fibre. This centring fibre will not go all the way through the bundle to CRIRES+ but will be transmitted to a local avalanche photo-diode (APD).

\subsection{Optical design}
\label{sec:optical_design}

\begin{figure}[p]
  \centering
  \includegraphics[width=0.95\textwidth]{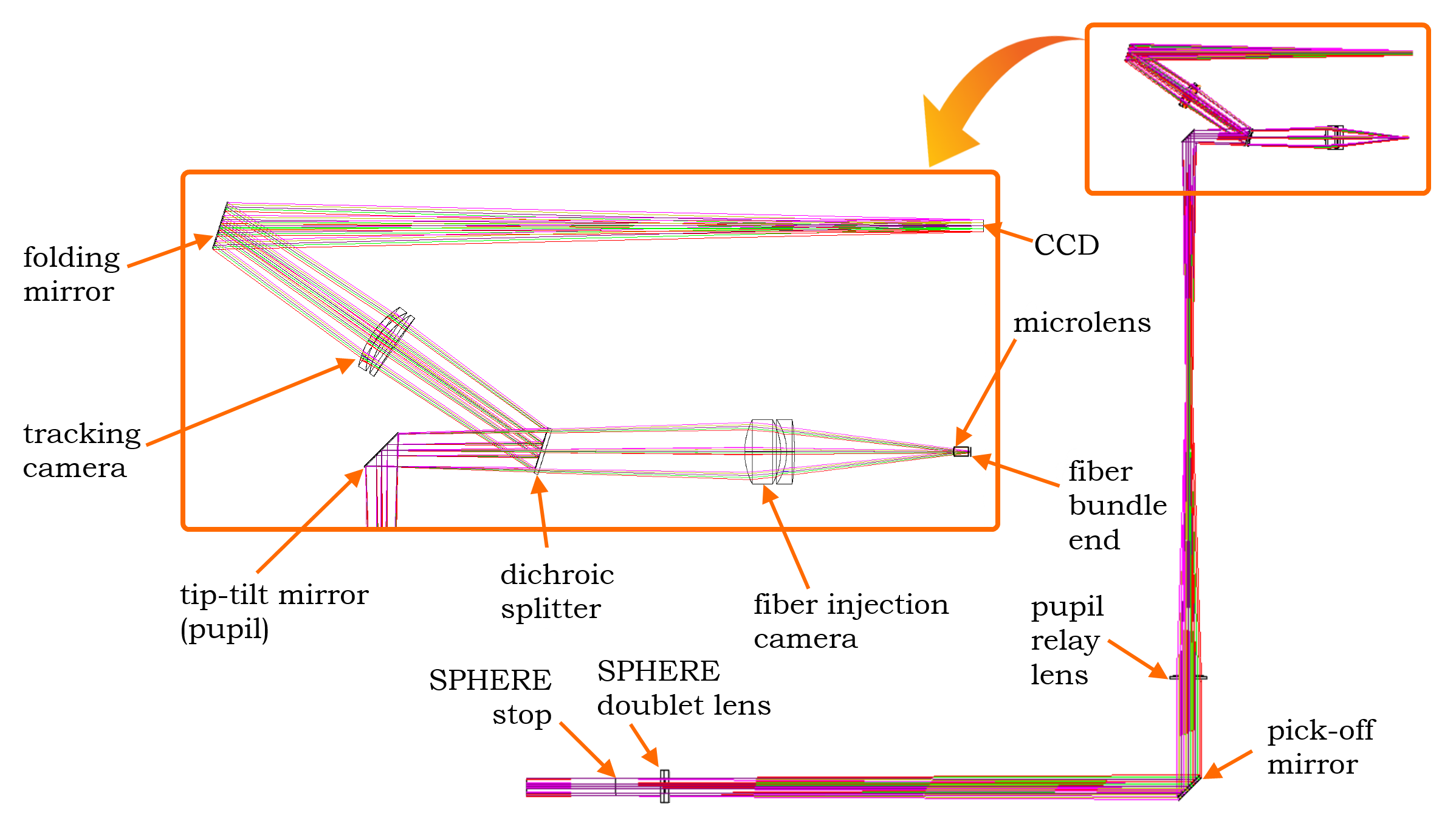}
  \caption{General optical layout of the SPHERE fibre injection unit. The pick-off mirror in the lower right corner corresponds to the retractable mirror described in Section~\ref{sec:location_conceptual_design}. An additional folding mirror in the tracking camera arm has been added with respect to the conceptual sketch presented in Figure~\ref{fig:hirise_fiu_concept}. The spot diagrams for both arms are presented in Figure~\ref{fig:hirise_optics_spots}.}
  \label{fig:hirise_optics_layout}
\end{figure}

\begin{figure}[p]
    \centering
    \hfill
    \begin{subfigure}[b]{0.36\textwidth}
        \includegraphics[width=\textwidth]{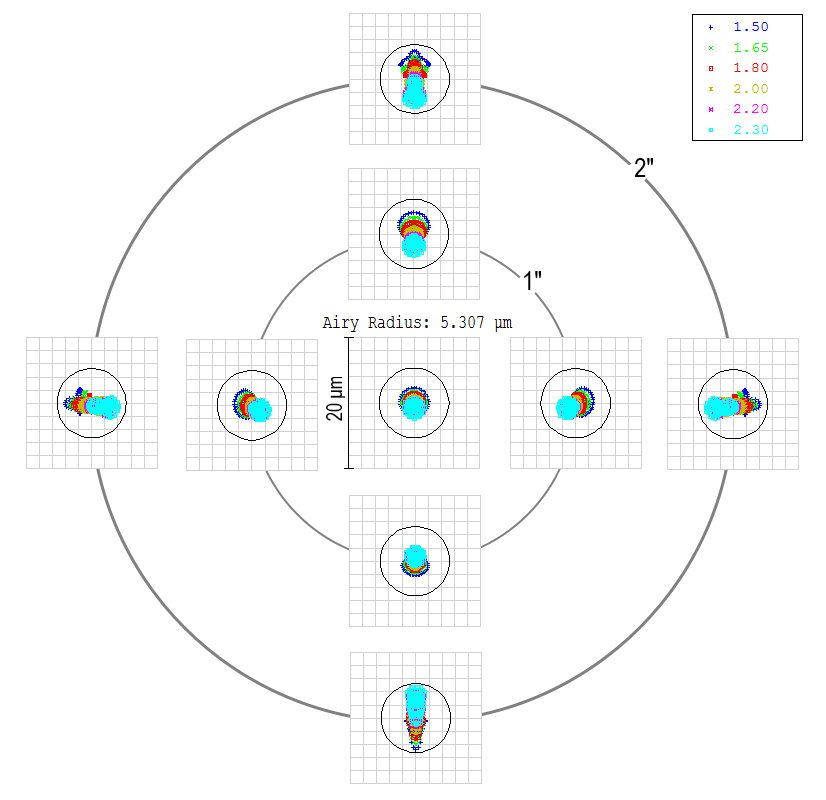}
        \caption{Fibre injection arm}
    \end{subfigure}
    \hfill
    \begin{subfigure}[b]{0.36\textwidth}
        \includegraphics[width=\textwidth]{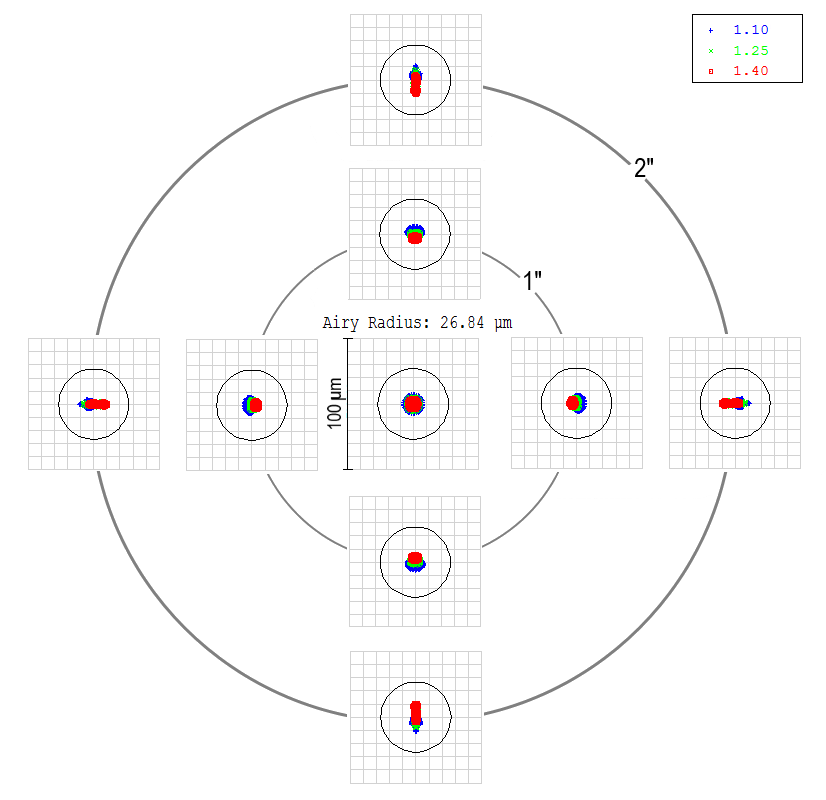}
        \caption{Tracking camera arm}
    \end{subfigure}\vspace{0.5em}
    \hfill~
    \caption{Spot diagrams for the fibre injection arm (left) and the tracking camera arm (right) of the fibre injection unit. The design is diffraction-limited at all positions in the field (2\as on-sky) over the full spectral range covered by each arm. The wavelength is indicated in microns in the upper right part of each plot.}
    \label{fig:hirise_optics_spots}
\end{figure}

The detailed optical design of the FIU is based on the latest version of the complete SPHERE optical design obtained at the end of the optical final design review in 2008. The general layout of the fibre injection arm and the tracking camera arm are presented in Figure~\ref{fig:hirise_optics_layout}, and some of the parameters are presented in Table~\ref{tab:optical_design}. The following points provide some additional information about the current design:

\begin{itemize}
    \setlength{\itemsep}{1pt}
    \setlength{\parskip}{0pt}
    \setlength{\parsep}{0pt}
    \item the pick-off mirror is located 350 mm from the last SPHERE lens;
    \item the pupil relay lens is an  off-the-shelf plano-convex lens Thorlabs LA5714 made of $\mathrm{CaF}_2$;
    \item the tip-tilt mirror clear aperture is 8$\times$11.3 mm because of the linear magnification;
    \item the dichroic beam splitter tilt angle is 17.5 deg;
    \item the tracking camera lens is a customized separated doublet lens (FTM16+$\mathrm{CaF}_2$);
    \item the injection optics are a customized separated doublet lens (FTM16+$\mathrm{CaF}_2$) followed by a customized microlens made of Schott IRG24 glass, which is placed 0.5 mm away from the fiber end. The focal plane is tilted by 0.5$^{\mathrm{o}}$ to compensate for the dichroic tilt.
\end{itemize}

The FIU is diffraction-limited in both the fibre injection and tracking camera arms over the appropriate wavelength ranges, as demonstrated by the spot diagrams in Figure~\ref{fig:hirise_optics_spots}. One of the difficulties identified in the design is the very short focal distance ($<$30 mm) between the injection doublet lens and the microlens, as well as between the microlens and the fibre bundle. These parameters are set by the numerical aperture of the fibres and therefore cannot be further optimised. Another difficulty is the accurate positioning of the first pick-off mirror: because of the long focal distance of the pupil relay lens, an accurate positioning of the pick-off mirror will be necessary to ensure that the pupil image properly falls on the tip-tilt mirror.

\begin{table}
    \caption{Optical design parameters}
    \label{tab:optical_design}
    \centering
    \begin{tabular}{lcc}
    \hline\hline
                         & Tracking camera & Fibre injection \\
    \hline         
    Spectral range       & 1.1--1.4~\mic   & 1.5--2.3~\mic   \\
    F/\#                 & 20              & 2.9             \\
    Linear field-of-view & $\pm$1.58 mm    & $\pm$0.23 mm    \\
    On-sky field-of-view    & $\pm$2.0\as     & $\pm$2.0\as     \\
    Final focal length   & 204.75 mm       & 29.44 mm        \\
    \hline
    \end{tabular} 
\end{table}

\subsection{Proposed operational scheme}
\label{sec:operational_scheme}

\begin{figure}
  \centering
  \includegraphics[width=0.6\textwidth]{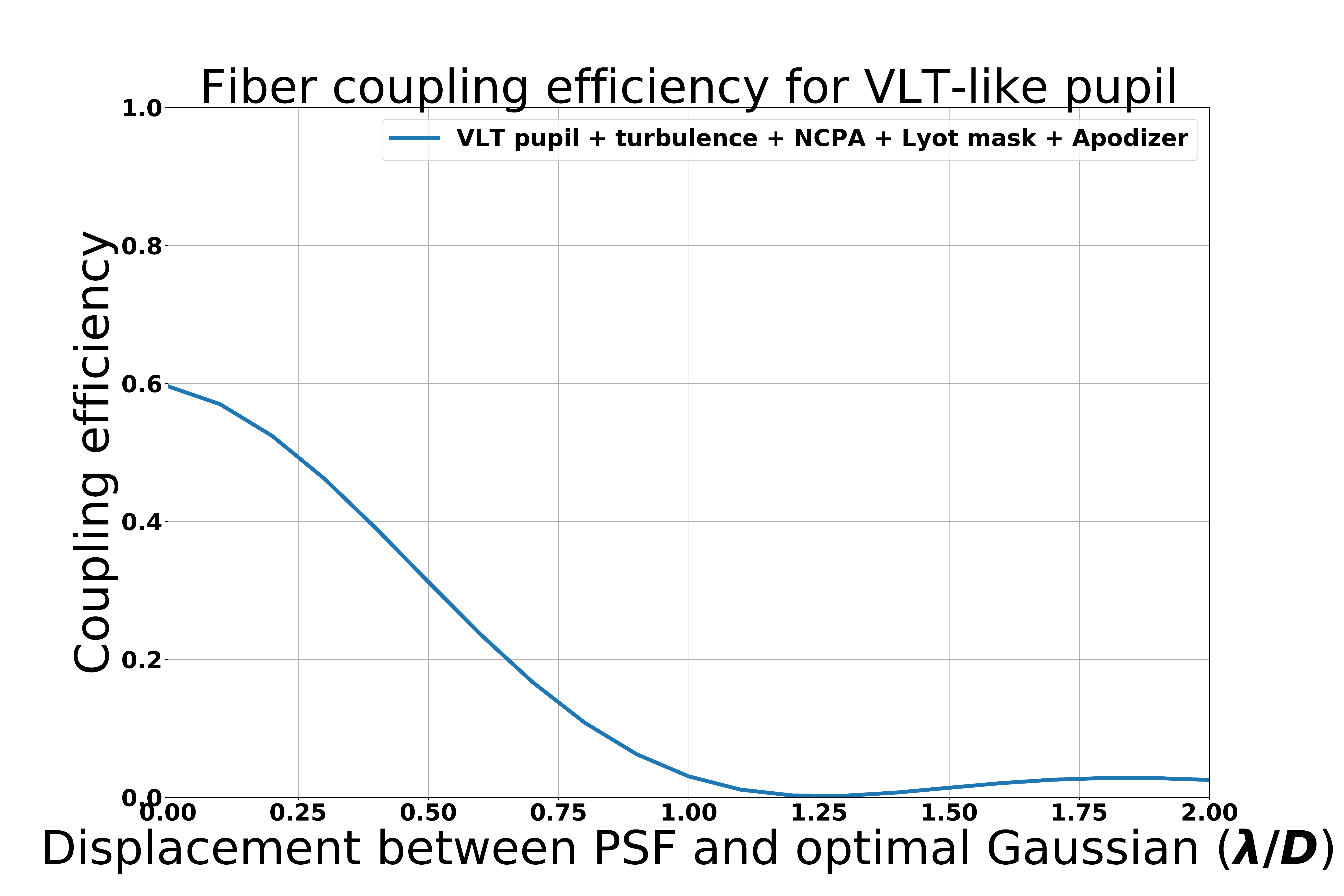}
  \caption{Coupling efficiency of the SPHERE apodised PSF in a single-mode fibre as a function of a lateral displacement (i.e. a tip or tilt error) in $\lambda/D$.}
  \label{fig:coupling efficiency}
\end{figure}

Besides fitting a new complex opto-mechanical system into an existing instrument, one of the challenges of HiRISE is to maximise the number of photons from the planet that are transmitted to the spectrograph, or in other words to maximise the coupling efficiency of the planetary signal into the science fibre. Previous studies have already explored how the coupling efficiency is impacted by the different features of the pupil\cite{Jovanovic2017}, such as the central obscuration or the spiders. The mismatch between the transmitted mode in the fibre (close to Gaussian) and the apodised VLT PSF already implies that only 60\% of the energy can be coupled into the fibre in a perfectly aligned system, but additional tip/tilt errors can further decrease the coupling efficiency, as demonstrated in Figure~\ref{fig:coupling efficiency}. This figure enables to estimate that to maintain the coupling efficiency $>$50\%, we have to center the planet PSF with an accuracy better than $\sim$$\lambda/(5D)$, which corresponds to $\sim$8~mas in $H$-band.

To reach this centring accuracy, we proposed a two-step calibration process based on the use of waffle spots\footnote{Sometimes referred to as satellite spots}. They are created by modulating the surface of the deformable with two sine waves in order to create replicas of the stellar PSF in the AO-corrected region that can be used for the determination of the PSF center\cite{Makidon2005,Marois2006b}. This procedure is regularly used in standard SPHERE coronagraphic observations and provide an accuracy of $\sim$1.2~mas\cite{Vigan2016a}. The first step of the centring procedure can be performed during daytime on the internal calibration point source. It consists in applying a waffle pattern on the DM, and then move the FIU tip-tilt mirror to place one of the waffle spots close to the centring fibre. Then, using gradient minimization algorithms, the tip-tilt mirror is moved in closed-loop so as to center the waffle spot precisely on the fibre, based on the flux coupled into the centring fibre measured by an APD. Then the procedure is repeated for the other 3 waffles spots, which in the end enables producing an interaction matrix that links the motion of the FIU tip-tilt mirror with the position of the image in the focal plane. 

The second part of the calibration is performed on-sky during night time. For this second part there are two options: either the companion is sufficiently bright so that the centring fibre can be centred directly on the companion, or it is too faint and the centring fibre has to be centred again on a waffle spot. In either case, the procedure ends with a \emph{blind offset} that moves the image so that the companion PSF falls on the science fibre. This offset will have been previously calibrated in the laboratory during integrations and is not expected to vary because is entirely depends on the manufactured geometry of the fibre bundle. Although the principle of this procedure is relatively simple, it has to be further studied to determine what is the expected accuracy that can be reached. In this scheme, the tracking camera is used to provide a visual feedback of the centring in closed-loop. 

In terms of SPHERE operations, the observations with the FIU should be completely independent of the rest of the instrument. The current baseline is to operate the instrument in normal \texttt{IRDIFS-EXT} mode in field-tracking, which means that the companion will remain at a fixed location during the whole observation. Although that will have an impact on the coronagraphic performance, because the Lyot stops are designed to be used in pupil-tracking observations, it will greatly simplify the observations. The option of operating in pupil-tracking remains however a possibility since the presence of the tracking camera could in principle enable implementing a closed-loop tracking algorithm. Further developments in this direction will be done in the coming months.


\section{CRIRES+}
\label{sec:crires+}

\subsection{Instrument description}

CRIRES+ is the successor of the Cryogenic Infrared Echelle Spectrograph (CRIRES) at the VLT. Formerly installed on UT1, the upgraded CRIRES+ instrument\cite{Dorn2016} will be hosted on the Nasmyth platform of UT3, entering operations in 2019. It is a high-resolution ($R=100\,000$), near-infrared ($0.95-5.4\mu$m) spectrograph, assisted by a MACAO\cite{Paufique2006} adaptive optics system. With the upgraded instrument, a new state-of the art NGC detector mosaic (three 2k$\times$2k HAWAII2-RG arrays) makes full use of the now cross-dispersed echelle format, boosting the simultaneous wavelength coverage by an order of magnitude compared with the old, single order CRIRES\cite{Kaeufl2004,Kaeufl2006,Kaeufl2008}.

Cross-dispersion is achieved with a set of six optimized gratings, one for each of the $YJHKLM$ bands. To ensure spectral purity and isolate the band, order sorting filters provide very good blocking in the out-of bands regions. The CRIRES+ spectrometer unit is inherited from old CRIRES with an R2 echelle grating. The echellogram hence covers 7-10 cross-dispersed orders depending on band, and about half ($Y$-band) to a quarter of the band ($K$-band) in one exposure. Consequently, 2-4 exposures are needed to record a full band without gaps in $Y$--$K$, respectively.

\begin{figure}
  \centering
  \includegraphics[width=1\textwidth]{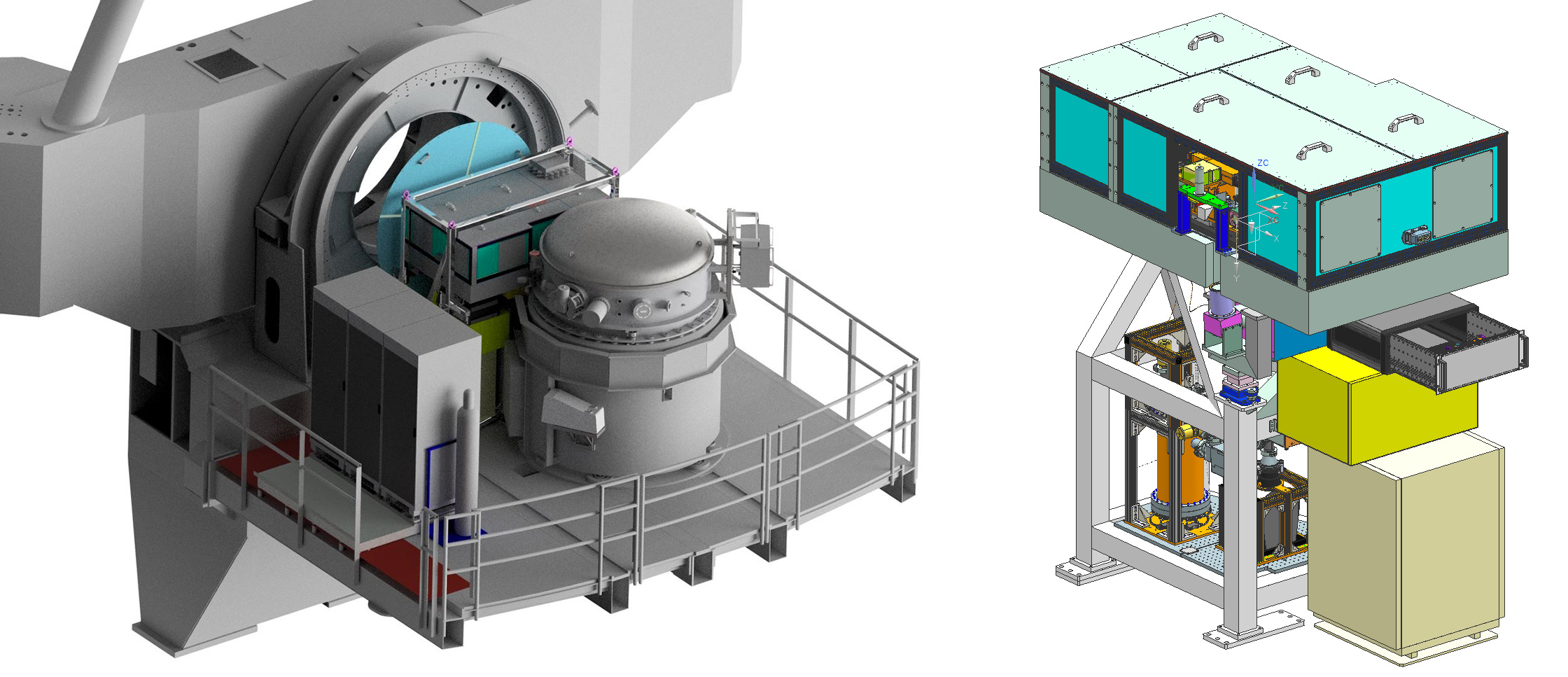}
  \caption{\emph{Left:} CRIRES+ on the Nasmyth platform of VLT/UT3. The spectrograph cryostat is readily visible, with the calibration unit and warm optics part in between to the telescope flange. \emph{Right:} The CRIRES+ calibration unit and adaptive optics (warm optics part). The structure is located in between the UT3 telescope adapter and the CRIRES+ instrument crysostat. The main components are an optical bench with the upper calibration unit and MACAO system, the lower calibration unit underneath the bench, and electronics cabinets and racks. The telescope beam enters at the center of the top part.}
  \label{fig:CR:UT3+CAL}
\end{figure}

CRIRES+ is composed of two main parts (see Figure~\ref{fig:CR:UT3+CAL}):

\begin{enumerate}
    \item The \emph{warm optics part} at ambient temperature that is mounted in front of the VLT telescope adapter. The warm part hosts the derotator, the AO system and pre-slit feed optics, and the calibration unit and associated feed and selection optics.
    Figure~\ref{fig:CR:UT3+CAL} right shows the opto-mechanical layout of the warm part. 
    
    \item The \emph{cold part} is a cryogenic vacuum vessel, interfaced with a dichroic entrance window. This part hosts the spectrograph's slit unit, the slit viewer camera and detector, the cross-disperser unit, and the spectrometer itself plus science detector system. The cold part is stabilized at 65K (including echelle grating) for thermal radiation suppression, while the detectors are operated as low as 35K for superior performance.
\end{enumerate}

\subsection{Calibration unit}
\label{sec:CR:calunit}

The CRIRES+ calibration unit\cite{Seemann2014} (CU) is mounted into the warm part, and comprises two main units. The lower CU provides emitting spectral sources and lamps, including a Fabry-Perot etalon system, for flat-fielding and wavelength calibration. The upper CU facilitates transmissive spectral calibrators (gas-cells as frequency standards and for precision RVs) as well as spatial calibrators for the AO and the spectrograph, a spectro-polarimetry unit, emission source, and reference fiber-injected light. The lower and upper CU can be combined, so that eg. the polarimeter, gas-cells, or spatial calibrators can be fed by light from the lower CU. Figure~\ref{fig:CR:CSU+PHU} shows a model view of the upper and lower CU.

\subsection{Fiber coupling to CRIRES+}
\label{sec:CR:fiber2CR}

\begin{figure}
  \centering
  \includegraphics[width=1\textwidth]{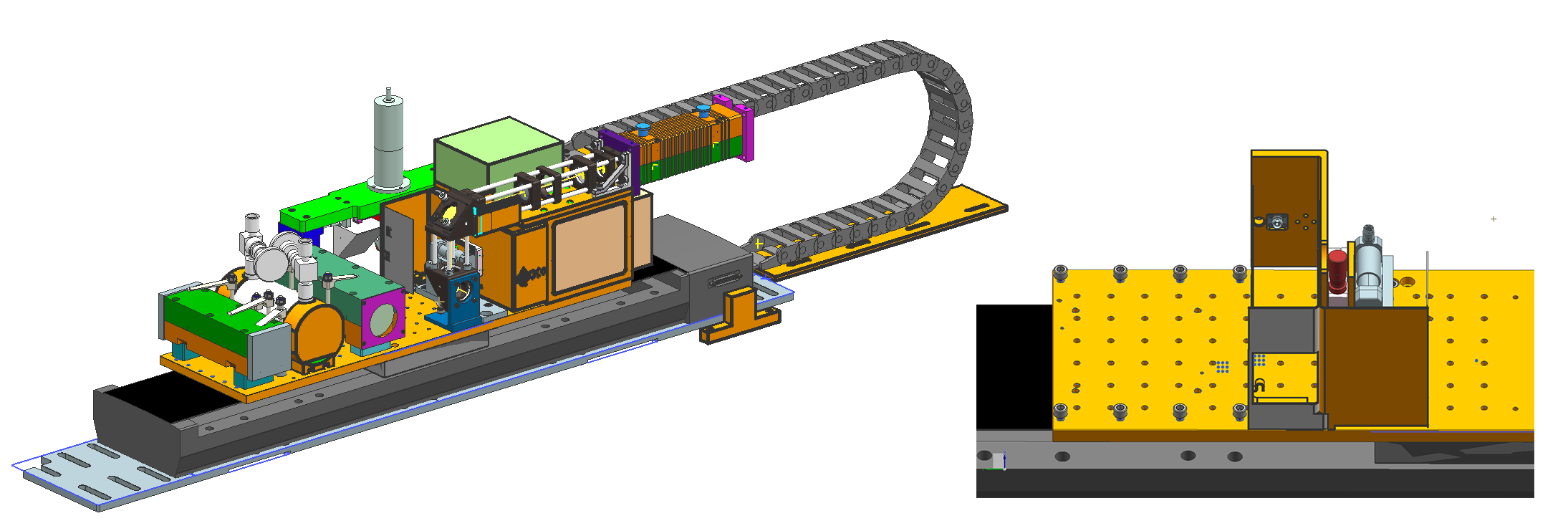}
  \caption{\emph{Left:} The CRIRES+ upper calibration unit serves to insert reference gas-cells (front left modules), spatial calibrators and a spectro-polarimetry unit (two rightmost modules), and to inject spectral reference light (middle module). Light from the VLT enters from behind, the VLT focal plane resides on the modules' front faces. Modules can be swapped and replaced, a HiRISE fiber injection module can be installed e.g. in place of a gas-cell. \emph{Right:} Simplified view of the available CRIRES+ injection unit for reference fibers, primarily used to calibrate MACAO and to provide spatial references. The fibers mount on XYZ-translation stages (additional rotational degree of freedom not shown). The unit is readily available to feed SPHERE fiber-coupled light in standard jack formats.}
  \label{fig:CR:CSU+PHU}
\end{figure}

The upper CU\cite{Seemann2014} hosts all devices on a movable precision stage, where the devices are grouped into modules that mount into opto-mechanically compatible slots (Figure~\ref{fig:CR:CSU+PHU}). These modules can be arranged more or less freely into the available slots, can be swapped, and taken out if needed. In this way, eg. a gas-cell module can be replaced by another one, and reinserted at any time at the exact location, thanks to reference pins.

A natural interface for optical fibers is the spatial calibrator module\cite{Seemann2018}. This unit is readily equipped with standard fiber connectors (FC/PC and SMA) on a movable support stage, and are designed to inject reference light into the telescope focal plane (Figure~\ref{fig:CR:CSU+PHU}). Three dynamic degrees of freedom (motorized focus along the optical axis, translation in the azimuth plane, and rotation) are available, plus manual control of translation in the altitude plane. Together, this allows for arbitrary fibre positioning within the field of view. In this way, a fiber or geometrically arranged fibres in standard jacks from SPHERE can be conveniently aligned to the spectrograph.

For a more complex injection setup or more control options on the fiber injection conditions, a designated module can be realized that fits into any of the upper CU unit slots. e.g. a fiber bundle with individual fibers rearranged into a linear array can be pre-configured such that all fibers follow the spectrograph slit along its full length. As the upper CU is a moving unit, any cabling and fibers need to be routed appropriately and stress-free. The CRIRES+ CU is equipped with a designated cable guide and wrap unit, which delivers all fiber cables at a constant physical routing (i.e. bending radius, stress) independent of the motion status of the upper CU.

SPHERE light is coupled into CRIRES+ via the HiRISE fibres in the same way as is any calibration or starlight, i.e. the light is injected into the object plane of the MACAO adaptive optics system. In this way, visible light ($400-800$\,nm) can serve to optimize the wavefront of the infrared light that is injected into the spectrograph, and hence optimize and stabilize the throughput on the entrance slit. The natural bandpass from SPHERE arriving at CRIRES+ via fiber contains no visible light, so that the MACAO system receives no signal. Consequently, in this case it can only run in open loop and no adaptive wavefront measurement and correction can be obtained, so that an accurate and temporally stable MACAO calibration is mandatory (essentially resembling a ``flat'' deformable mirror). Such calibration can be derived with the CRIRES+ CU, or on the HiRISE fibers if also visible reference light is injected on the SPHERE side.

\section{Simulations}
\label{sec:simulations}

To simulate the expected performance of the combination of SPHERE and CRIRES+ with the proposed fiber coupling unit we developed an end-to-end model for the integrated system. This model takes spectra for both the planet and the star, and then takes into account various wavelength-dependent transmission losses, fiber coupling efficiency, planet-to-star contrast and sources of noise that are picked up in the system. The resultant spectrum is then processed by a basic data reduction pipeline in order to estimate the SNR of specific atmospheric species after cross-correlation with template spectra for each of the species. The model is also used to assess which parts of the system can be optimized to improve the SNR.

\subsection{Simulation model and assumptions}
\label{sec:model_assumptions}

The spectral models and spectral templates are generated using the 1D radiative-convective equilibrium code ATMO \cite{Tremblin2015} to compute a grid of self-consistent pressure-temperature profiles and chemical equilibrium abundances for a range of effective temperatures and gravities. The line-by-line radiative transfer calculation in ATMO is then performed using these profiles and abundances as inputs to generate a high resolution thermal emission spectrum. Then the spectral templates are generated using a similar method to that described in Wang et al. (2017)\cite{Wang2017}. The template for a given molecular absorber is the emission spectrum calculated by removing all opacities in the model atmosphere apart from the respective absorber and the absorption from $\mathrm{H_2}$-$\mathrm{H_2}$ and $\mathrm{H_2}$-$\mathrm{He}$ collisions. This allows the contributions to the thermal emission of dominant molecular absorbers such as $\mathrm{H_2O}$, $\mathrm{CO}$ and $\mathrm{CH_4}$ to be isolated. In total, ATMO contains 20 molecular and atomic opacity sources primarily originating from the ExoMol database\cite{Tennyson2016}, and has been most recently described in Goyal et al. (2018)\cite{Goyal2018}.

For the stellar spectrum we start with a BT-NextGen (AGSS2009)\cite{Allard2012} \footnote{To be be replaced by a higher resolution PHOENIX model \cite{Husser2013}} model of a host star, as they best fit the observed stellar SEDs. This model is interpolated to a regular grid with a wavelength step size of 1 angstrom from 600 nm to 30 microns. Photometric SED data points in $J$, $H$ and $K_{s}$ are taken for the star from VOSA 6.0 \cite{Bayo2008}. After integrating over each of the bands with the corresponding filter response curve we perform a least square fit to obtain a flux scaling factor to later rescale the models to realistic values. This flux rescaling factor is in the same order of magnitude as a geometric flux scaling factor derived from the stellar radius $d_\mathrm{radius}$ and distance $d_\mathrm{distance}$, namely $(d_\mathrm{radius}/d_\mathrm{distance})^2$. To obtain the spectrum of the star we interpolate the original spectrum to a spectrum resolution of $R$, convert the units to photons and rescale with the flux scaling parameter. For the planet we take the previously described model from ATMO at 1 angstrom resolution and the planet's known delta magnitude with respect to the star in a certain band ($H$-band). Using the filter curve of the corresponding band we integrate the spectrum of both the planet and the star and define a scaling factor that would rescale the planetary spectrum to be $10^{\Delta mag/2.5}$ times fainter than the star in that band. To obtain the spectrum of the planet we again interpolate the original planetary spectrum to the spectral resolution, convert it to photons and rescale with the scaling factor.

In the next steps we modulate the spectra with the atmospheric and instrumental transmission and add background. The first step is to take into account the atmospheric transmission. For this we use the ESO SkyCalc online client \cite{Noll2012,Jones2013} configured with $\mathrm{PWV} = 3.5 \mathrm{mm}$, airmass of 1, average season, average time of day, with contributions from moonlight, starlight, zodiacal light, lower and upper atmosphere, airglow and thermal background included. In addition we included a $300 \mathrm{K}$ instrumental background component with an emissivity of 0.5.  

\begin{figure}
  \centering
  \includegraphics[width=0.9\textwidth]{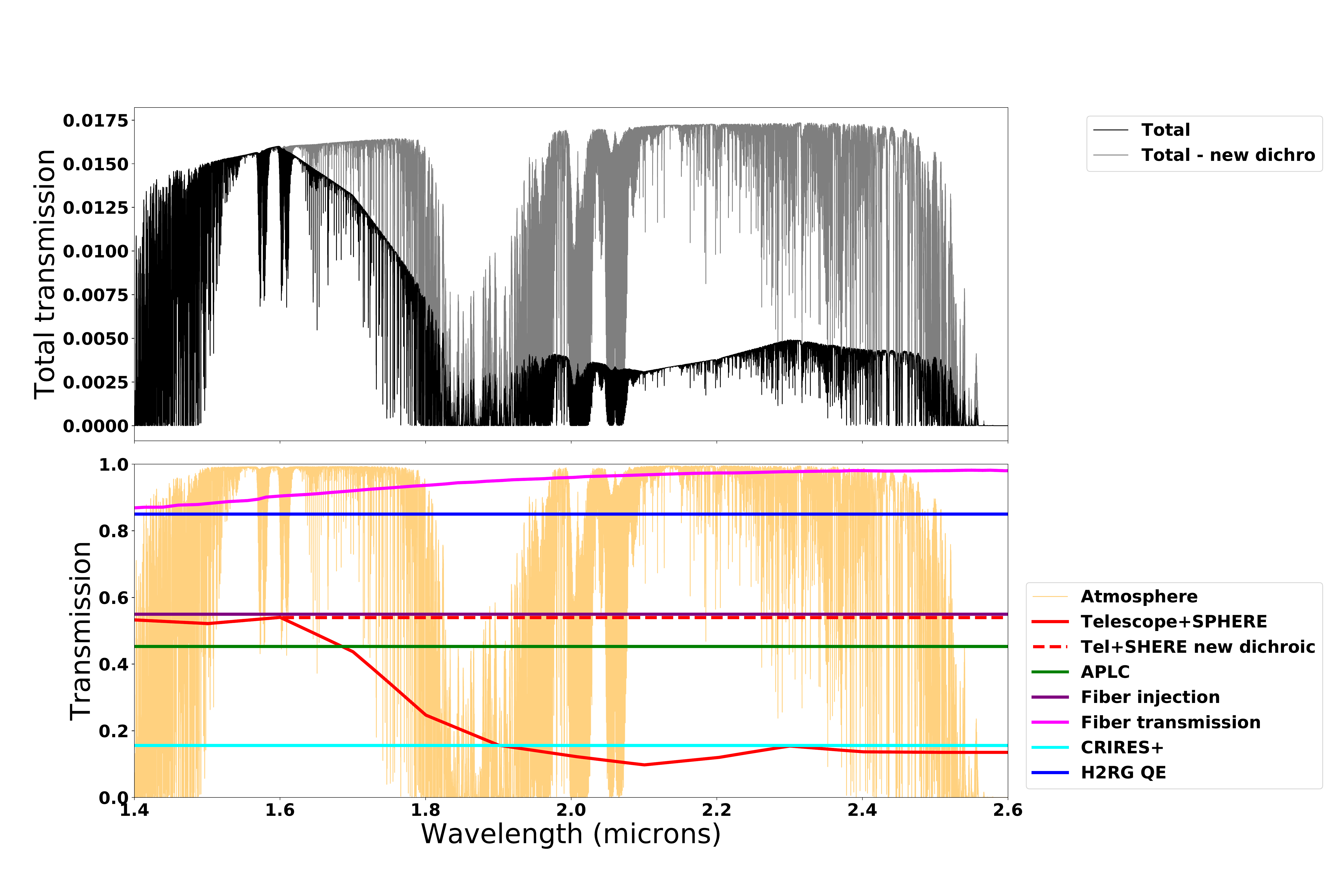}
  \caption{Expected HiRISE total transmission, decomposed per component. Depending on the efficiency of the dicroic, the total end-to-end transmission is between 0.5 and 1.5\%.}
  \label{fig:hirise_transmission}
\end{figure}

The telescope transmission and the SPHERE CPI values are taken from measurements done during the assembly and integration phase. In addition, we add the transmission loss of the APLC apodizer and Lyot mask. Regarding the fiber we need to take into account several aspects, most importantly the fiber coupling efficiency and the fiber transmission. The fiber coupling efficiency is determined by a calculation following the theory described in Wagner et al. (1982)\cite{Wagner1982} and Jovanovic et al. (2017)\cite{Jovanovic2017}. We use recent measurements of the SPHERE amplitude and NCPA errors, residual AO turbulence and the pupil masks of the coronagraph to generate a PSF that is then matched with an optimally sized Gaussian beam. For these initial calculations we assume that the fiber mode field diameter scales directly with the instrumental PSF. We use a constant 55\% fiber coupling efficiency, considering an imperfect alignment on the order of $\lambda/(10D)$. The relation between coupling efficiency and tip-tilt error is given in Figure \ref{fig:coupling efficiency}. The apodizer increases the coupling efficiency by a few percent but blocks a large fraction of light. It may in fact be better for the end-to-end performance to remove the apodizer and accept a reduced contrast for increased throughput. A more detailed field and wavelength dependent coupling efficiency, including influence of the coronagraph will be determined in the near future. The fiber transmission is derived from manufacturer specifications\footnote{~\url{http://leverrefluore.com/}} for a single-mode ZBLAN fiber and converted to a 55 meter length fiber. The CRIRES+ transmission is set to 16\% following reports from integration, and finally the detector quantum efficiency is set to 85\% following the CRIFORS\footnote{~\url{https://github.com/ivh/crifors}} simulator default settings. In addition to the planetary and stellar light we also take the radiance component from the ESO SkyCalc models and propagate it through our optical system. We take into account the surface area of the telescope, exposure time, the width of the spectral bins and (in the case of the background) the angular extent of the fiber on the sky. To simulate the stellar signal going into the fiber we use the stellar flux and multiply with the strength of the halo. In this case we assume a flat $5 \times 10^{-5}$. Even for a strongly stellar halo dominated system such as $\beta$ Pictoris a halo strength of $10^{-4}$ is enough to suppress the stellar halo to a level lower than the other noise sources. The transmission contributions used in this work are summarized in Table~\ref{tab:transmission_budget} and Figure~\ref{fig:hirise_transmission}.

\begin{figure}
  \centering
  \includegraphics[width=0.8\textwidth]{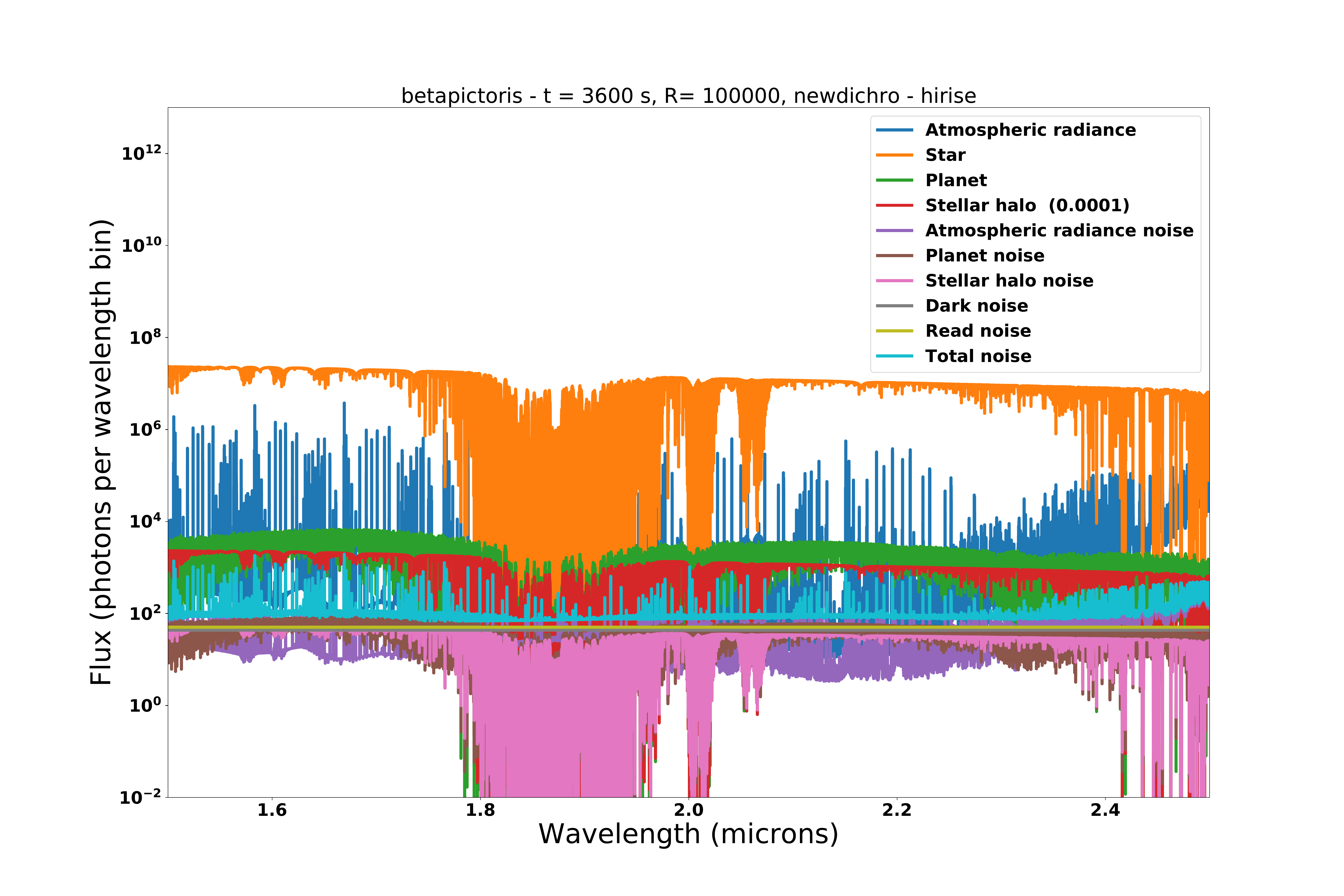}
  \caption{Simulated spectrum of Beta Pic like system within 1 hour of exposure time.}
  \label{fig:hirise_betapic_spectrum}
\end{figure}

\begin{table}
    \caption{Transmission budget}
    \label{tab:transmission_budget}
    \centering
    \begin{tabular}{lcc}
    \hline \hline
    Element                   & \multicolumn{2}{c}{Transmission} \\
                              & \emph{H}-band & \emph{K}-band    \\
    \hline      
    Atmosphere                & 96\% & 94\% \\
    SPHERE CPI \& optics      & 25\%     &  7\% \\
    APLC, apodiser            & \multicolumn{2}{c}{57\%} \\
    APLC, Lyot stop           & \multicolumn{2}{c}{79\%} \\
    Fibre coupling efficiency & \multicolumn{2}{c}{55\%} \\
    Fibre                     & 90\% & 97\% \\
    CRIRES+                   & \multicolumn{2}{c}{16\%} \\
    \hline
    Total                     & 0.9\% & 0.3\% \\
    \hline
    \end{tabular} 
\end{table}

Shot noise is generated and injected for the different additive flux components (stellar halo, planet and atmospheric radiance). In addition, we inject read noise (7 electrons per readout for optimal Fowler sampling) and dark noise (0.01 electrons per second) per pixel. The spectral resolution elements stretch out in a worst case scenario to a height of 15 pixels and a width of approximately 2 pixels on the orders falling on the detector at the shortest wavelengths in $K$-band. The detector noise sources are scaled with the square-root of the amount of pixels per resolution element. An example spectrum and its components are shown in Figure~\ref{fig:hirise_betapic_spectrum}. Note that the height along the spatial dimension can be reduced as the fiber extraction unit can be designed to come in with a different F-number, although a too large mismatch may lead to additional transmission losses within CRIRES+. Saturation can be limited by taking a sequence of shorter exposures if necessary.

\subsection{Preliminary results}
\label{sec:preliminary_results}

Using the earlier described model, we generate a spectrum of stellar systems resembling $\beta$ Pictoris b, 51 Eridani b and HIP\,65426 b in terms of star-to-planet contrast and brightness. For the planet we use an object of spectral type L5 with a \teff of 1600K and $\logg=4.0$, which is a good match for $\beta$ Pic b and HIP\,65426 b, but admittedly not for 51 Eri b. A larger grid of models in terms of \teff and \logg will be generated and explored in the future.

Figure~\ref{fig:hirise_betapic_spectrum} already shows that with the improved stellar suppression, the stellar halo is no longer the dominating noise factor, as is the case with brighter systems \cite{Snellen2015}. Instead, the total noise is dominated by the shot noise on the planet.

\begin{figure}[t]
  \centering
  \includegraphics[width=0.6\textwidth]{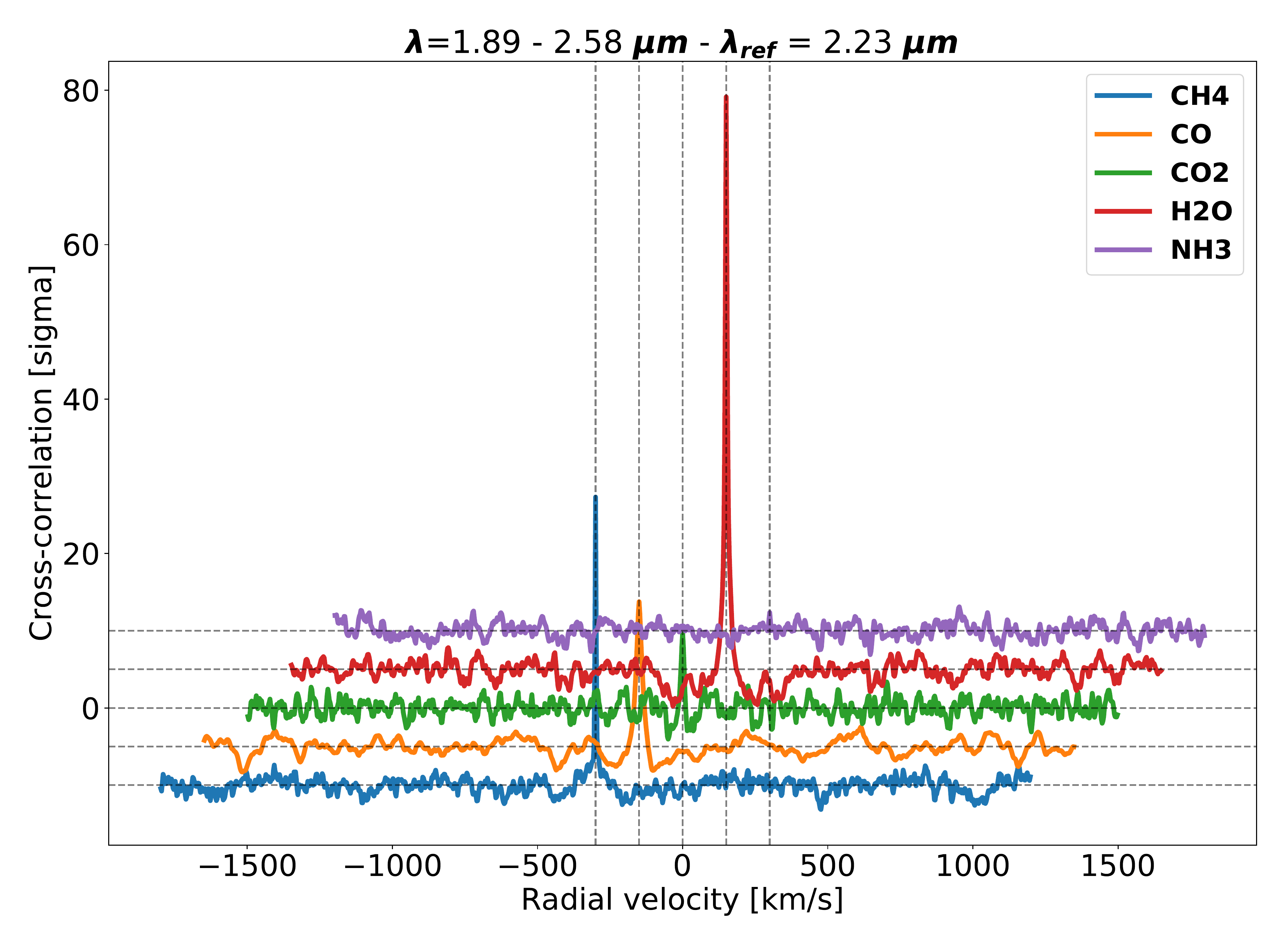}
  \caption{Example of cross-correlation functions for multiple atmospheric species over $K$-band.}
  \label{fig:hirise_betapic_ccf}
\end{figure}

In order to derive the SNRs of the detection of individual atmospheric species we run a cross-correlation of the planet signal with the templates for individual species. For these initial simulations we assume that we can perfectly subtract the stellar and atmospheric contributions from the total flux signal, leaving only the planet light with all the noise contributions. We remove the continuum of the spectrum by running a FFT high-pass filter. For each individual atmospheric species we also have isolated the spectral template and removed its continuum. The flattened planetary signal is then cross-correlated with the spectral templates with velocities from -1500 to 1500~km/s. For an hour of observing time on $\beta$ Pictoris~b the cross-correlation signal is clearly detected for most of the species (see also Figure~\ref{fig:hirise_betapic_ccf}). Unfortunately, no variance in time can be used to determine the signal to noise of the CCF as we expect to have only a few long integrations (to reduce the impact of the read noise). The signal to noise is therefore calculated by dividing the peak of the CCF with the noise at large positive and negative velocities. This yields a crude SNR value directly from the CCF function but in some cases the noise can be dominated by cross-correlation structure (self-similarity of the spectral templates) using this technique. Alternative SNR estimation methods that are less sensitive to this noise factor will be explored in future work. The detection of species with smaller strengths (in particular $\mathrm{NH_3}$) appears to be strongly limited by the CCF noise and a different approach should be taken to derive a more representative SNR, not influenced by the self-similarity in the spectral templates.

\begin{figure}[t]
  \centering
  \includegraphics[width=0.9\textwidth]{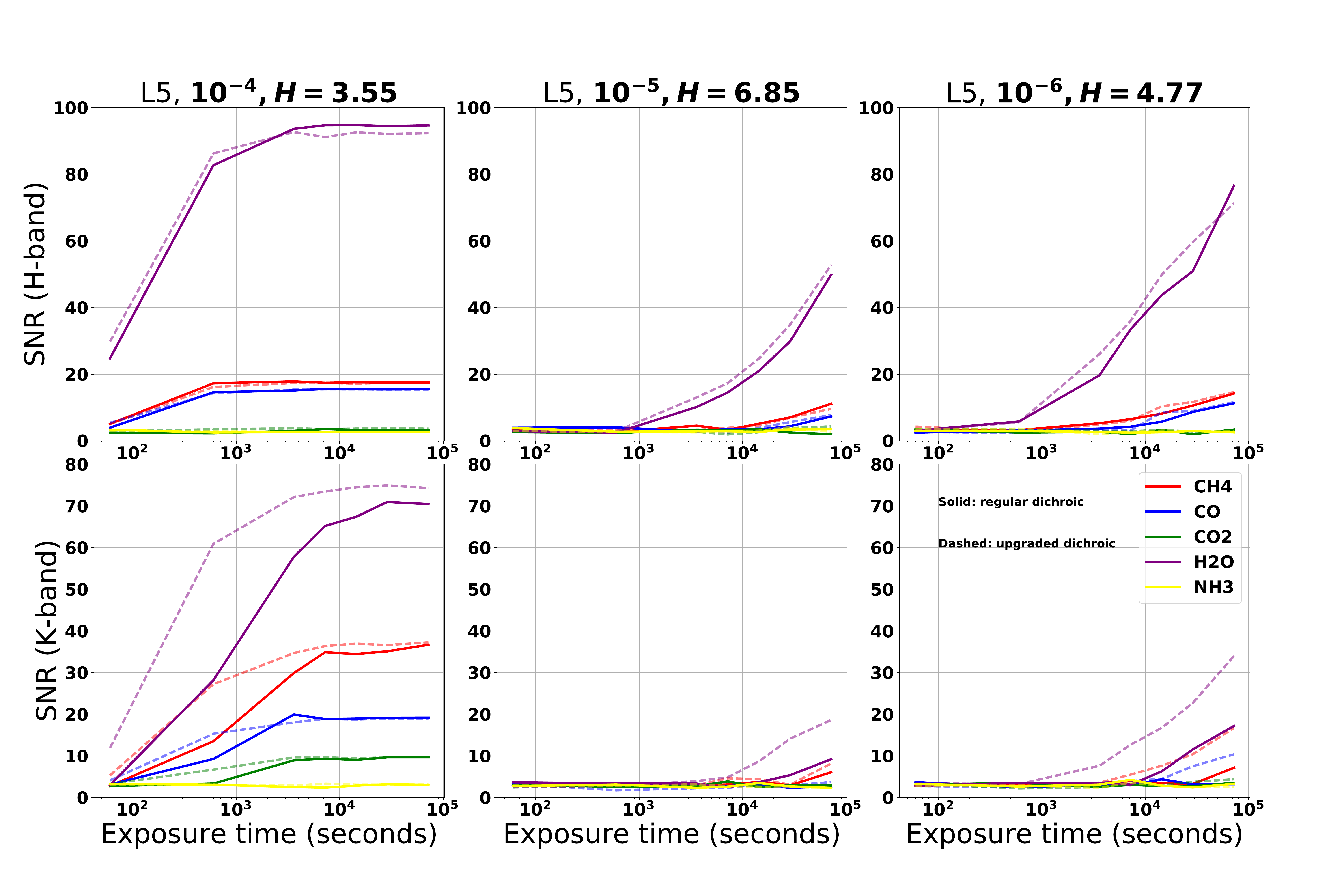}
  \caption{Signal to noise of CCF versus exposure time. For a $\beta$ Pictoris b-like system CCF noise limits are reached after 1 to 2 hours of observing time. In the case of a HIP\,65426 b-like and 51 Eri b-like systems, increased exposure times still lead to increased SNR.}
  \label{fig:hirise_snr_time}
\end{figure}

To investigate the impact of using a dichroic with higher throughput in $K$-band we ran again the simulations with the CPI transmission given a flat dicroic throughput at wavelengths beyond 1.6~\mic. Figure~\ref{fig:hirise_snr_time} shows the SNR dependence on exposure time for each of the atmospheric species and the three simulated stellar systems in the full $H$- and $K$-band. The simulated dataset with the old dichroic is shown with solid lines and the upgraded dichroic performance is shown with dashed lines. As the dichroic performance drops off at 1.6~\mic the impact on the $H$-band performance is limited but the improvement is very clearly seen in $K$-band, almost doubling the signal to noise. For the three simulated stellar systems significant $\mathrm{H}_2\mathrm{O}$ detections are made with a night of observing time.

\section{Conclusions \& prospects}
\label{sec:conclusion_prospects}

We have presented the status of project HiRISE at the ESO Very Large Telescope, which aims at implementing a fibre injection coupling between the SPHERE exoplanet imager and the CRIRES+ high-resolution spectrograph. This coupling would enable a step forward in the characterisation of known, young, giant exoplanets that have been directly imaged until now.

The general implementation is based on three elements. The first is a fibre injection unit that will be implemented in SPHERE. It is designed as a completely independent opto-mechanical system located in the IFS arm of SPHERE, where it will benefit from the full infrastructure of the instrument: stable environment, high-quality optics, high-order AO correction, very fine stabilisation of the tip-tilt, and full-fledged coronagraph. It will include a retractable pick-off mirror that will be inserted into the beam when the FIU is in use. The IFU includes two arms, one for the fibre injection in $H$- and $K$-band, and the other with a tracking camera in $J$-band. For the operations, the most challenging part is the accurate centring of the planet PSF on the science fibre. For this step we foresee the use of waffle spots created in the focal plane by modulating the deformable mirror of the instrument and a dedicated centring fibre connected to an APD. A calibration procedure will enable to calibrate the motion of the FIU tip-tilt mirror, so that the science fibre can be moved precisely at any accessible location in the focal plane.

The second element is the bundle of single-mode fibres. While the exact geometry of the bundle is still being defined, the path that the bundle will have to follow at the telescope has been explored. The most suitable option is similar to the one adopted by the VLT/FLAMES instrument, where the fibre bundle will go from the SPHERE Nasmyth platform down beneath the telescope azimuth floor and then back up on the other Nasmyth platform where the CRIRES+ instrument will be located. The total length of fibre in this configuration is of the order of 55~meters.

The third element is the opto-mechanical interface in CRIRES+. CRIRES+ is currently being upgrade at ESO in Garching to become a cross-dispersed echelle spectograph equipped with 3 new Hawaii-2RG detectors. The upgraded instrument will be much more efficient and will provide a much wider spectral coverage in a single observation than the original CRIRES+. At the entrance of CRIRES+ is a calibration stage that includes gas cells dedicated to the wavelength calibration, a calibration lamp, a polarimetric calibration unit and calibration fibres. This calibration stage is the most natural location to implement a fibre bundle coming from SPHERE. The opto-mechanical interface will be explored in the coming months.

Finally, we have presented the first performance estimations of the HiRISE system. By implementing the coupling of SPHERE with CRIRES+ our simulations show that for bright stellar systems similar to Beta Pictoris, which are normally limited in high-dispersion spectroscopy studies by the shot noise on the stellar halo, we see a strong reduction of this noise source and the SNR on the atmospheric species becomes limited by the shot noise on the planet signal. Adding a $H$- and $K$-band optimized dichroic to SPHERE roughly quadruples the throughput beyond 1.6 microns which would give a doubling of the SNR for detections in that band. Increasing total throughput of the system and, in addition, reducing the impact of the readnoise and dark noise by concentrating the light onto fewer pixels on the spectrograph is critical to achieving a similar improvement for fainter targets, which are otherwise limited by the CRIRES+ detector noises.

The design of HiRISE is on-going, with a preliminary timeline of $\sim$1.5 years to reach a final design and hopefully an implementation and first science in the subsequent months. The HiRISE system is meant both as a new observing mode for SPHERE and as a pathfinder for future instrumentation. HiRISE will open new scientific possibilities that are currently not accessible to SPHERE or CRIRES+ on their own, for a small sample of known objects (10-15). On the longer term, HiRISE is an important pathfinder experiment because coupling high-contrast imaging and high-dispersion spectroscopy may be one of the few realistic possibilities that will enable the detection of Earth-like planets in the coming decades.

\acknowledgments

This project has received funding from the European Research Council (ERC) under the European Union's Horizon 2020 research and innovation programme, grant agreements No. 757561 (HiRISE) \& 678777 (ICARUS).

\smallskip

SPHERE is an instrument designed and built by a consortium consisting of IPAG (Grenoble, France), MPIA (Heidelberg, Germany), LAM (Marseille, France), LESIA (Paris, France), Laboratoire Lagrange (Nice, France), INAF - Osservatorio di Padova (Italy), Observatoire de Gen\`eve (Switzerland), ETH Zurich (Switzerland), NOVA (Netherlands), ONERA (France) and ASTRON (Netherlands) in collaboration with ESO. SPHERE was funded by ESO, with additional contributions from CNRS (France), MPIA (Germany), INAF (Italy), FINES (Switzerland) and NOVA (Netherlands). SPHERE also received funding from the European Commission Sixth and Seventh Framework Programmes as part of the Optical Infrared Coordination Network for Astronomy (OPTICON) under grant number RII3-Ct-2004-001566 for FP6 (2004-2008), grant number 226604 for FP7 (2009-2012) and grant number 312430 for FP7 (2013-2016).

\smallskip

This publication makes use of VOSA, developed under the Spanish Virtual Observatory project supported from the Spanish MINECO through grant AyA2017-84089.

\bibliography{paper}

\begin{thebibliography}{10}

\bibitem{Jovanovic2015}
{Jovanovic}, {Martinache}, {Guyon}, {et~al.}, ``{The Subaru Coronagraphic
  Extreme Adaptive Optics System: Enabling High- Contrast Imaging on
  Solar-System Scales},'' {\em Publications of the Astronomical Society of the
  Pacific}~{\bf 127},  890 (Sept. 2015).

\bibitem{Macintosh2014}
{Macintosh}, {Graham}, {Ingraham}, {et~al.}, ``{First light of the Gemini
  Planet Imager},'' {\em Proceedings of the National Academy of Science}~{\bf
  111},  12661--12666 (Sept. 2014).

\bibitem{Beuzit2008}
{Beuzit}, {Feldt}, {Dohlen}, {et~al.}, ``{SPHERE: a 'Planet Finder' instrument
  for the VLT},'' in [{\em Ground-based and Airborne Instrumentation for
  Astronomy II}{\nolinebreak\hspace{0.1em}]},   {\bf 7014},  701418 (July
  2008).

\bibitem{Macintosh2015}
{Macintosh}, {Graham}, {Barman}, {et~al.}, ``{Discovery and spectroscopy of the
  young jovian planet 51 Eri b with the Gemini Planet Imager},'' {\em
  Science}~{\bf 350},  64--67 (Oct. 2015).

\bibitem{Chauvin2017}
{Chauvin}, {Desidera}, {Lagrange}, {et~al.}, ``{Discovery of a warm, dusty
  giant planet around HIP 65426},'' {\em \aap}~{\bf 605},  L9 (Sept. 2017).

\bibitem{Zurlo2016}
{Zurlo}, {Vigan}, {Galicher}, {et~al.}, ``{First light of the VLT planet finder
  SPHERE. III. New spectrophotometry and astrometry of the HR 8799 exoplanetary
  system},'' {\em \aap}~{\bf 587},  A57 (Mar. 2016).

\bibitem{Vigan2016a}
{Vigan}, {Bonnefoy}, {Ginski}, {et~al.}, ``{First light of the VLT planet
  finder SPHERE. I. Detection and characterization of the substellar companion
  GJ 758 B},'' {\em \aap}~{\bf 587},  A55 (Mar. 2016).

\bibitem{Samland2017}
{Samland}, {Molli{\`e}re}, {Bonnefoy}, {et~al.}, ``{Spectral and atmospheric
  characterization of 51 Eridani b using VLT/SPHERE},'' {\em \aap}~{\bf 603},
  A57 (July 2017).

\bibitem{Chilcote2017}
{Chilcote}, {Pueyo}, {De Rosa}, {et~al.}, ``{1-2.4 {\ensuremath{\mu}}m Near-IR
  Spectrum of the Giant Planet {\ensuremath{\beta}} Pictoris b Obtained with
  the Gemini Planet Imager},'' {\em \aj}~{\bf 153},  182 (Apr. 2017).

\bibitem{Greenbaum2018}
{Greenbaum}, {Pueyo}, {Ruffio}, {et~al.}, ``{GPI Spectra of HR 8799 c, d, and e
  from 1.5 to 2.4 {\ensuremath{\mu}}m with KLIP Forward Modeling},'' {\em
  \aj}~{\bf 155},  226 (June 2018).

\bibitem{Soummer2005}
{Soummer}, ``{Apodized Pupil Lyot Coronagraphs for Arbitrary Telescope
  Apertures},'' {\em \apj}~{\bf 618},  L161--L164 (Jan. 2005).

\bibitem{Boccaletti2008}
{Boccaletti}, {Abe}, {Baudrand}, {et~al.}, ``{Prototyping coronagraphs for
  exoplanet characterization with SPHERE},'' in [{\em Adaptive Optics
  Systems}{\nolinebreak\hspace{0.1em}]},  {\em \procspie} {\bf 7015},  70151B
  (July 2008).

\bibitem{Guyon2014}
{Guyon}, {Hinz}, {Cady}, {Belikov}, \& {Martinache}, ``{High Performance Lyot
  and PIAA Coronagraphy for Arbitrarily Shaped Telescope Apertures},'' {\em
  \apj}~{\bf 780},  171 (Jan. 2014).

\bibitem{Soummer2007}
{Soummer}, {Ferrari}, {Aime}, \& {Jolissaint}, ``{Speckle Noise and Dynamic
  Range in Coronagraphic Images},'' {\em \apj}~{\bf 669},  642--656 (Nov.
  2007).

\bibitem{Vigan2016b}
{Vigan}, {N'Diaye}, {Dohlen}, {et~al.}, ``{Stop-less Lyot coronagraph for
  exoplanet characterization: first on-sky validation in VLT/SPHERE},'' in
  [{\em Advances in Optical and Mechanical Technologies for Telescopes and
  Instrumentation II}{\nolinebreak\hspace{0.1em}]},   {\bf 9912},  991226 (July
  2016).

\bibitem{Racine1999}
{Racine}, {Walker}, {Nadeau}, {Doyon}, \& {Marois}, ``{Speckle Noise and the
  Detection of Faint Companions},'' {\em Publications of the Astronomical
  Society of the Pacific}~{\bf 111},  587--594 (May 1999).

\bibitem{Marois2006a}
{Marois}, {Lafreni{\`e}re}, {Doyon}, {Macintosh}, \& {Nadeau}, ``{Angular
  Differential Imaging: A Powerful High-Contrast Imaging Technique},'' {\em
  \apj}~{\bf 641},  556--564 (Apr. 2006).

\bibitem{Lafreniere2007}
{Lafreni{\`e}re}, {Doyon}, {Nadeau}, {et~al.}, ``{Improving the Speckle Noise
  Attenuation of Simultaneous Spectral Differential Imaging with a Focal Plane
  Holographic Diffuser},'' {\em \apj}~{\bf 661},  1208--1217 (June 2007).

\bibitem{Mugnier2009}
{Mugnier}, {Cornia}, {Sauvage}, {et~al.}, ``{Optimal method for exoplanet
  detection by angular differential imaging},'' {\em Journal of the Optical
  Society of America A}~{\bf 26},  1326 (May 2009).

\bibitem{Soummer2012}
{Soummer}, {Pueyo}, \& {Larkin}, ``{Detection and Characterization of
  Exoplanets and Disks Using Projections on Karhunen-Lo{\`e}ve Eigenimages},''
  {\em \apj}~{\bf 755},  L28 (Aug. 2012).

\bibitem{Cantalloube2015}
{Cantalloube}, {Mouillet}, {Mugnier}, {et~al.}, ``{Direct exoplanet detection
  and characterization using the ANDROMEDA method: Performance on VLT/NaCo
  data},'' {\em \aap}~{\bf 582},  A89 (Oct. 2015).

\bibitem{Wang2016}
{Wang}, {Graham}, {Pueyo}, {et~al.}, ``{The Orbit and Transit Prospects for
  {\ensuremath{\beta}} Pictoris b Constrained with One Milliarcsecond
  Astrometry},'' {\em \aj}~{\bf 152},  97 (Oct. 2016).

\bibitem{Nielsen2017}
{Nielsen}, {De Rosa}, {Rameau}, {et~al.}, ``{Evidence That the Directly Imaged
  Planet HD 131399 Ab Is a Background Star},'' {\em \aj}~{\bf 154},  218 (Dec.
  2017).

\bibitem{Wertz2017}
{Wertz}, {Absil}, {G{\'o}mez Gonz{\'a}lez}, {et~al.}, ``{VLT/SPHERE robust
  astrometry of the HR8799 planets at milliarcsecond- level accuracy. Orbital
  architecture analysis with PyAstrOFit},'' {\em \aap}~{\bf 598},  A83 (Feb.
  2017).

\bibitem{Bonnefoy2016}
{Bonnefoy}, {Zurlo}, {Baudino}, {et~al.}, ``{First light of the VLT planet
  finder SPHERE. IV. Physical and chemical properties of the planets around
  HR8799},'' {\em \aap}~{\bf 587},  A58 (Mar. 2016).

\bibitem{Snellen2010}
{Snellen}, {de Kok}, {de Mooij}, \& {Albrecht}, ``{The orbital motion, absolute
  mass and high-altitude winds of exoplanet HD209458b},'' {\em \nat}~{\bf 465},
   1049--1051 (June 2010).

\bibitem{Crossfield2012}
{Crossfield}, {Hansen}, \& {Barman}, ``{Ground-based, Near-infrared
  Exospectroscopy. II. Tentative Detection of Emission from the Extremely Hot
  Jupiter WASP-12b},'' {\em \apj}~{\bf 746},  46 (Feb. 2012).

\bibitem{Martins2015}
{Martins}, {Santos}, {Figueira}, {et~al.}, ``{Evidence for a spectroscopic
  direct detection of reflected light from <ASTROBJ>51 Pegasi b</ASTROBJ>},''
  {\em \aap}~{\bf 576},  A134 (Apr. 2015).

\bibitem{Hoeijmakers2018}
{Hoeijmakers}, {Snellen}, \& {van Terwisga}, ``{Searching for reflected light
  from {\ensuremath{\tau}} Bootis b with high-resolution ground-based
  spectroscopy: Approaching the 10<SUP>-5</SUP> contrast barrier},'' {\em
  \aap}~{\bf 610},  A47 (Feb. 2018).

\bibitem{Sparks2002}
{Sparks} \& {Ford}, ``{Imaging Spectroscopy for Extrasolar Planet Detection},''
  {\em \apj}~{\bf 578},  543--564 (Oct. 2002).

\bibitem{Riaud2007}
{Riaud} \& {Schneider}, ``{Improving Earth-like planets' detection with an ELT:
  the differential radial velocity experiment},'' {\em \aap}~{\bf 469},
  355--361 (July 2007).

\bibitem{Snellen2015}
{Snellen}, {de Kok}, {Birkby}, {et~al.}, ``{Combining high-dispersion
  spectroscopy with high contrast imaging: Probing rocky planets around our
  nearest neighbors},'' {\em \aap}~{\bf 576},  A59 (Apr. 2015).

\bibitem{Lagrange2010}
{Lagrange}, {Bonnefoy}, {Chauvin}, {et~al.}, ``{A Giant Planet Imaged in the
  Disk of the Young Star {\ensuremath{\beta}} Pictoris},'' {\em Science}~{\bf
  329},  57 (July 2010).

\bibitem{Konopacky2013}
{Konopacky}, {Barman}, {Macintosh}, \& {Marois}, ``{Detection of Carbon
  Monoxide and Water Absorption Lines in an Exoplanet Atmosphere},'' {\em
  Science}~{\bf 339},  1398--1401 (Mar. 2013).

\bibitem{Mawet2016}
{Mawet}, {Wizinowich}, {Dekany}, {et~al.}, ``{Keck Planet Imager and
  Characterizer: concept and phased implementation},'' in [{\em Adaptive Optics
  Systems V}{\nolinebreak\hspace{0.1em}]},   {\bf 9909},  99090D (July 2016).

\bibitem{Wang2017}
{Wang}, {Mawet}, {Ruane}, {Hu}, \& {Benneke}, ``{Observing Exoplanets with High
  Dispersion Coronagraphy. I. The Scientific Potential of Current and
  Next-generation Large Ground and Space Telescopes},'' {\em \aj}~{\bf 153},
  183 (Apr. 2017).

\bibitem{Mawet2017}
{Mawet}, {Ruane}, {Xuan}, {et~al.}, ``{Observing Exoplanets with
  High-dispersion Coronagraphy. II. Demonstration of an Active Single-mode
  Fiber Injection Unit},'' {\em \apj}~{\bf 838},  92 (Apr. 2017).

\bibitem{Lovis2017}
{Lovis}, {Snellen}, {Mouillet}, {et~al.}, ``{Atmospheric characterization of
  Proxima b by coupling the SPHERE high-contrast imager to the ESPRESSO
  spectrograph},'' {\em \aap}~{\bf 599},  A16 (Mar. 2017).

\bibitem{Pasquini2002}
{Pasquini}, {Avila}, {Blecha}, {et~al.}, ``{Installation and commissioning of
  FLAMES, the VLT Multifibre Facility},'' {\em The Messenger}~{\bf 110},  1--9
  (Dec. 2002).

\bibitem{Dorn2016}
{Dorn}, {Follert}, {Bristow}, {et~al.}, ``{The ''+'' for CRIRES: enabling
  better science at infrared wavelength and high spectral resolution at the ESO
  VLT},'' in [{\em Ground-based and Airborne Instrumentation for Astronomy
  VI}{\nolinebreak\hspace{0.1em}]},   {\bf 9908},  99080I (Aug. 2016).

\bibitem{Hugot2012}
{Hugot}, {Ferrari}, {El Hadi}, {et~al.}, ``{Active optics methods for exoplanet
  direct imaging. Stress polishing of supersmooth aspherics for VLT-SPHERE
  planet finder},'' {\em \aap}~{\bf 538},  A139 (Feb. 2012).

\bibitem{Fusco2006}
{Fusco}, {Rousset}, {Sauvage}, {et~al.}, ``{High-order adaptive optics
  requirements for direct detection of extrasolar planets: Application to the
  SPHERE instrument},'' {\em Optics Express}~{\bf 14},  7515 (Jan. 2006).

\bibitem{Petit2014}
{Petit}, {Sauvage}, {Fusco}, {et~al.}, ``{SPHERE eXtreme AO control scheme:
  final performance assessment and on sky validation of the first auto-tuned
  LQG based operational system},'' in [{\em Adaptive Optics Systems
  IV}{\nolinebreak\hspace{0.1em}]},   {\bf 9148},  91480O (Aug. 2014).

\bibitem{Sauvage2014}
{Sauvage}, {Fusco}, {Petit}, {et~al.}, ``{Wave-front sensor strategies for
  SPHERE: first on-sky results and future improvements},'' in [{\em SPIE
  Conference Series}{\nolinebreak\hspace{0.1em}]},   {\bf 9148} (Aug. 2014).

\bibitem{Thalmann2008}
{Thalmann}, {Schmid}, {Boccaletti}, {et~al.}, ``{SPHERE ZIMPOL: overview and
  performance simulation},'' in [{\em Ground-based and Airborne Instrumentation
  for Astronomy II}{\nolinebreak\hspace{0.1em}]},  {\em \procspie} {\bf 7014},
  70143F (July 2008).

\bibitem{Claudi2008}
{Claudi}, {Turatto}, {Gratton}, {et~al.}, ``{SPHERE IFS: the spectro
  differential imager of the VLT for exoplanets search},'' in [{\em
  Ground-based and Airborne Instrumentation for Astronomy
  II}{\nolinebreak\hspace{0.1em}]},  {\em \procspie} {\bf 7014},  70143E (July
  2008).

\bibitem{Dohlen2008}
{Dohlen}, {Langlois}, {Saisse}, {et~al.}, ``{The infra-red dual imaging and
  spectrograph for SPHERE: design and performance},'' in [{\em Ground-based and
  Airborne Instrumentation for Astronomy II}{\nolinebreak\hspace{0.1em}]},
  {\em \procspie} {\bf 7014},  70143L (July 2008).

\bibitem{Vigan2018}
{Vigan}, {N'Diaye}, {Dohlen}, \& {et~al.}, ``{On-sky compensation of non-common
  path aberrations with the ZELDA wavefront sensor in VLT/SPHERE},'' {\em
  \procspie} {\bf 10703} (2018).

\bibitem{Baudoz2010}
{Baudoz}, {Dorn}, {Lizon}, {et~al.}, ``{The differential tip-tilt sensor of
  SPHERE},'' in [{\em Ground-based and Airborne Instrumentation for Astronomy
  III}{\nolinebreak\hspace{0.1em}]},   {\bf 7735},  77355B (July 2010).

\bibitem{Jovanovic2017}
{Jovanovic}, {Schwab}, {Guyon}, {et~al.}, ``{Efficient injection from large
  telescopes into single-mode fibres: Enabling the era of ultra-precision
  astronomy},'' {\em \aap}~{\bf 604},  A122 (Aug. 2017).

\bibitem{Makidon2005}
{Makidon}, {Sivaramakrishnan}, {Perrin}, {et~al.}, ``{An Analysis of
  Fundamental Waffle Mode in Early AEOS Adaptive Optics Images},'' {\em
  Publications of the Astronomical Society of the Pacific}~{\bf 117},  831--846
  (Aug. 2005).

\bibitem{Marois2006b}
{Marois}, {Lafreni{\`e}re}, {Macintosh}, \& {Doyon}, ``{Accurate Astrometry and
  Photometry of Saturated and Coronagraphic Point Spread Functions},'' {\em
  \apj}~{\bf 647},  612--619 (Aug. 2006).

\bibitem{Paufique2006}
{Paufique}, {Biereichel}, {Donaldson}, {et~al.}, ``{MACAO-CRIRES, a step
  towards high-resolution spectroscopy},'' {\em ArXiv Astrophysics e-prints}
  (Aug. 2006).

\bibitem{Kaeufl2004}
{K{\"a}ufl}, {Ballester}, {Biereichel}, {et~al.}, ``{CRIRES: a high-resolution
  infrared spectrograph for ESO's VLT},'' in [{\em Society of Photo-Optical
  Instrumentation Engineers (SPIE) Conference
  Series}{\nolinebreak\hspace{0.1em}]},  {Moorwood} \& {Iye}, eds., {\em
  Society of Photo-Optical Instrumentation Engineers (SPIE) Conference Series}
  {\bf 5492},  1218--1227 (Sept. 2004).

\bibitem{Kaeufl2006}
{K{\"a}ufl}, {Amico}, {Ballester}, {et~al.}, ``{Good Vibrations: Report from
  the Commissioning of CRIRES},'' {\em The Messenger}~{\bf 126},  32--36 (Dec.
  2006).

\bibitem{Kaeufl2008}
{K{\"a}ufl}, {Amico}, {Ballester}, {et~al.}, ``{CRIRES: commissioning and first
  science results},'' in [{\em Society of Photo-Optical Instrumentation
  Engineers (SPIE) Conference Series}{\nolinebreak\hspace{0.1em}]},  {\em
  Society of Photo-Optical Instrumentation Engineers (SPIE) Conference Series}
  {\bf 7014} (Aug. 2008).

\bibitem{Seemann2014}
{Seemann}, {Anglada-Escude}, {Baade}, {et~al.}, ``{Wavelength calibration from
  1-5{$\mu$}m for the CRIRES+ high-resolution spectrograph at the VLT},'' in
  [{\em Ground-based and Airborne Instrumentation for Astronomy
  V}{\nolinebreak\hspace{0.1em}]},  {\em \procspie} {\bf 9147},  91475G (Aug.
  2014).

\bibitem{Seemann2018}
{Seemann}, {Anwand-Heerwart}, H., \& Valenti, ``{Laboratory and system
  performance of the VLT/CRIRES+ infra-red Fabry-Perot Etalon calibrator},'' in
  [{\em Advances in Optical and Mechanical Technologies for Telescopes and
  Instrumentation III}{\nolinebreak\hspace{0.1em}]},  {\em \procspie} {\bf
  10705},  10706 (Aug. 2018).

\bibitem{Tremblin2015}
{Tremblin}, {Amundsen}, {Mourier}, {et~al.}, ``{Fingering Convection and
  Cloudless Models for Cool Brown Dwarf Atmospheres},'' {\em The Astrophysical
  Journal Letters}~{\bf 804},  L17 (May 2015).

\bibitem{Tennyson2016}
{Tennyson}, {Yurchenko}, {Al-Refaie}, {et~al.}, ``{The ExoMol database:
  Molecular line lists for exoplanet and other hot atmospheres},'' {\em Journal
  of Molecular Spectroscopy}~{\bf 327},  73--94 (Sept. 2016).

\bibitem{Goyal2018}
{Goyal}, {Mayne}, {Sing}, {et~al.}, ``{A library of ATMO forward model
  transmission spectra for hot Jupiter exoplanets},'' {\em Monthly Notices of
  the Royal Astronomical Society}~{\bf 474},  5158--5185 (Mar. 2018).

\bibitem{Allard2012}
{Allard}, {Homeier}, \& {Freytag}, ``{Models of very-low-mass stars, brown
  dwarfs and exoplanets},'' {\em Philosophical Transactions of the Royal
  Society of London Series A}~{\bf 370},  2765--2777 (June 2012).

\bibitem{Husser2013}
{Husser}, {Wende-von Berg}, {Dreizler}, {et~al.}, ``{A new extensive library of
  PHOENIX stellar atmospheres and synthetic spectra},'' {\em \aap}~{\bf 553},
  A6 (May 2013).

\bibitem{Bayo2008}
{Bayo}, {Rodrigo}, {Barrado Y Navascu{\'e}s}, {et~al.}, ``{VOSA: virtual
  observatory SED analyzer. An application to the Collinder 69 open cluster},''
  {\em \aap}~{\bf 492},  277--287 (Dec. 2008).

\bibitem{Noll2012}
{Noll}, {Kausch}, {Barden}, {et~al.}, ``{An atmospheric radiation model for
  Cerro Paranal. I. The optical spectral range},'' {\em \aap}~{\bf 543},  A92
  (July 2012).

\bibitem{Jones2013}
{Jones}, {Noll}, {Kausch}, {Szyszka}, \& {Kimeswenger}, ``{An advanced
  scattered moonlight model for Cerro Paranal},'' {\em \aap}~{\bf 560},  A91
  (Dec. 2013).

\bibitem{Wagner1982}
{Wagner} \& {Tomlinson}, ``{Coupling efficiency of optics in single-mode fiber
  components},'' {\em \ao}~{\bf 21},  2671--2688 (Aug. 1982).

\end{thebibliography}
\bibliographystyle{spiebib_short}

\end{document}